\documentclass[%
 reprint, amsmath,amssymb,aps,pra]{revtex4-2}

\usepackage{graphicx}
\usepackage{dcolumn}
\usepackage{bm}
\usepackage{braket}
\usepackage{hyperref}
\usepackage[mathlines]{lineno}

\begin{document}

\preprint{APS/123-QED}

\title{Anisotropy-Induced Absorption in a Large Zeeman Manifold.}

\author{Nayan Sharma}
\affiliation{
	Department of Physics, Sikkim University, 6th Mile Samdur, East Sikkim, India -737102
}
\author{Ranjit Kumar Singh}
\affiliation{
	Department of Physics, Sikkim University, 6th Mile Samdur, East Sikkim, India -737102
}
\author{Ajay Tripathi}
\email{ajay$\_$t$\_$2000@yahoo.com}
\affiliation{
	Department of Physics, Sikkim University, 6th Mile Samdur, East Sikkim, India -737102
}
\author{Souvik Chatterjee}
\affiliation{Department of Chemistry, Amity Institute of Applied Sciences,
	Amity University, Sector-125, Noida, Uttar Pradesh 201313, India.
}
\author{Prasanta K. Panigrahi}
\affiliation{
	Department of Physical Sciences, Indian Institute of Science Education
	and Research Kolkata, Mohanpur 741246, West Bengal, India.
}
\begin{abstract}
We investigate absorption induced by optical anisotropy at two-photon resonance using pump-probe spectroscopy on the D$_2$ line of $^{87}$Rb.  Our approach combines detailed 13-level and 16-level atomic models, capturing all relevant Zeeman sub-levels, with elliptically polarized pump and probe beams. The study reveals that, at two-photon resonance, the 13-level system exhibits symmetric conversion between transmission and absorption for both positive and negative ellipticity. In contrast, the 16-level model, which includes the $F'=1$ manifold, displays a loss of symmetry, with absorption observed only for positive ellipticity. These theoretical predictions are supported by experimental measurements. Our findings provide insight into how optical anisotropy at two-photon resonance can be engineered through atomic structure and light polarization.
\end{abstract}

\maketitle

\section{Introduction}
The interaction of light with multi-level atoms gives rise to a wide range of phenomena driven by quantum interference among different excitation pathways. Effects such as electromagnetically induced transparency (EIT)\cite{fleischhauer2005electromagnetically,finkelstein2023practical}, electromagnetically induced absorption (EIA)\cite{lezama1999electromagnetically,liao2020microwave}, coherent population trapping (CPT)\cite{alzetta1976experimental,wu2025coherent}, and various forms of wave mixing\cite{walker2012trans,mathew2021single} and two-photon absorption \cite{hendrickson2010observation,tabakaev2022spatial} are prominent examples, especially in alkali atomic vapors like Rb and Cs \cite{thaicharoen2024rydberg,pati2021synchronous,brazhnikov2019ultrahigh}. Typically, these phenomena are investigated using pump–probe configurations, where the pump field modifies the atomic state and the probe field measures the resulting changes in absorption or transmission. Precise control over these processes has enabled a variety of applications, including quantum memories \cite{ma2022high,lei2022electromagnetically}, slow light \cite{derose2023producing,chen2024triple}, precision magnetometry \cite{gonzalez2024sensitivity,andryushkov2022vector}, cooling techniques for atomic clocks \cite{krysenko2023ground}, and advanced quantum information processing \cite{khazali2023all,shao2024rydberg}.

\begin{figure}[htp!]
	\centering
	\includegraphics[width=0.5\textwidth]{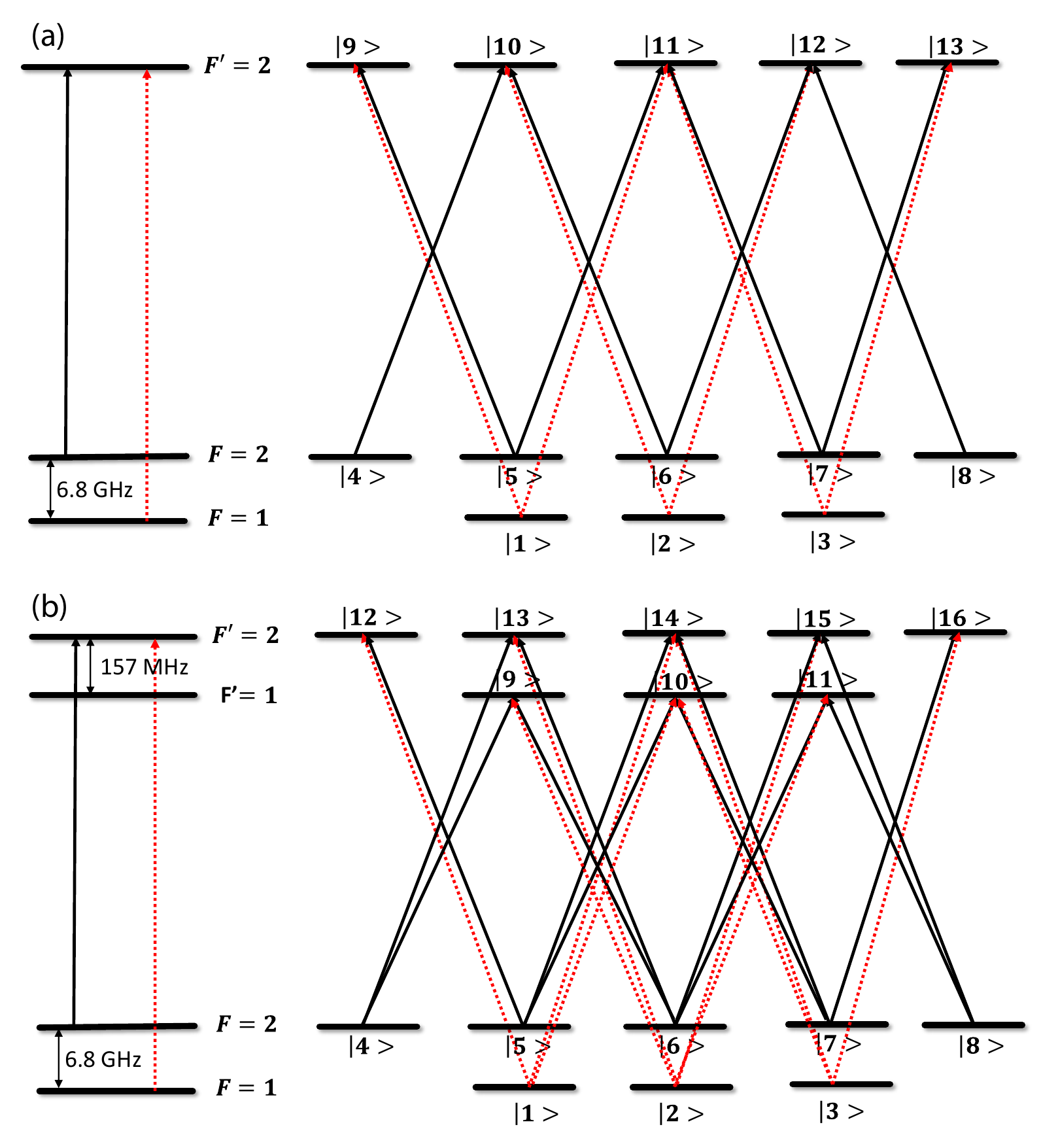}
	\caption{Model system:  (a) 13 - level model comprising of Zeeman sublevels of $\ket{F=1}$, $\ket{F=2}$ and $\ket{F'=2}$ states of D$_2$ line (b) 16 - level model including the Zeeman sublevels of $\ket{F'=1}$.Red (Dashed) represents weak probe beam and Black(solid) represents strong pump beam.}
	\label{fig1}
\end{figure}

In the absence of external magnetic fields, it has been demonstrated that resonance properties can be finely tuned by introducing additional light beams \cite{chanu2012conversion,singh2023light} or by controlling the relative phase between fields \cite{xu2013tuning,zhang2024controlling}. Degenerate atomic systems, in particular, show pronounced sensitivity to the polarization states of the applied light \cite{bhattarai2018tuning,ram2010effect,brazhnikov2011polarization}.
When a static magnetic field is applied, multi-level non-degenerate atomic systems exhibit an even richer variety of phenomena \cite{chauhan2021enhancement,sargsyan2012splitting,mishra2018electromagnetically,bhushan2019effect,ma2017paschen}. These include the simultaneous presence of dark (transmission) and bright (absorption) resonances, whose origins can differ fundamentally depending on the number of atomic levels involved and the magnetic field strength. Moreover, parameters such as magnetic field intensity and beam power are known to govern the conversion between transmission and absorption resonances \cite{singh2021competition,subba2020understanding,das2021effects,bhattarai2019study}. Additionally, the polarization of light and the influence of nearby hyperfine transitions play crucial roles in shaping the observed phenomena in real atomic systems like $^{87}$Rb \cite{hassan2025effect}.\\
In the case of a transverse magnetic field with D$_2$ line of $^{87}$ Rb, it was observed that some EIT resonances converted to absorption at around 18 G \cite{singh2021competition}, indicating the onset of significant influence from nearby hyperfine states on the resonance behavior. In this work, we investigate a similar scenario but in the presence of a longitudinal magnetic field, focusing on the effect of varying the ellipticity of the light beams. When linearly polarized light interacts with atoms, it generates what is known as pure alignment-a state where the populations of atoms are evenly distributed between opposite magnetic sublevels. This creates an anisotropic distribution of angular momentum without producing an overall magnetic moment. On the other hand, circularly polarized light leads to pure orientation, where atoms preferentially occupy one stretched magentic sublevel, resulting in a net magnetic moment aligned with the light's propagation direction. Between these two extremes, elliptically polarized light give rise to intermediate states called elliptical dark states (EDS). These states emerge through laser-induced coherences among ground-state Zeeman sublevels and embody the characteristics of both alignment and orientation, producing complex and asymmetric patterns. This phenomenon has been extensively described in studies like those by Milner et.al.~\cite{milner1998multilevel,milner1999arbitrary}.\\
We demonstrate that as light polarization changes from linear to pure circular, the intermediate anisotropic population distributions induced by elliptically polarized light give rise to optical anisotropy, resulting in a distinct conversion between EIT and absorption. To this end, we use two atomic models where we employ a comprehensive approach that incorporates all relevant atomic levels in our calculations. The first is a 13-level system, which includes all Zeeman sublevels of the ground state ($F=1$ and $F=2$) and the excited state ($F'=2$). In this configuration, the probe laser is locked to the $F=1 \rightarrow F'=2$ transition, while the pump laser is scanned across $F=2 \rightarrow F'=2$. To examine the influence of nearby hyperfine states, we extend this to a 16-level model by including all Zeeman sublevels of the $F'=1$ excited state. The polarization of the pump and probe fields is kept orthogonal to each other. The corresponding energy level diagrams are shown in Fig.~\ref{fig1}(a) and (b).\\
Numerical calculations with the 13-level model (excluding nearby hyperfine interactions) show that the conversion from transmission to absorption occurs for the 13-level model without the influence of nearby hyperfine states, and is symmetrically distributed for both positive and negative ellipticity. However, when the 16-level model is used to account for adjacent hyperfine states, this conversion is observed only for positive ellipticity, and the symmetry is lost demonstrating the influence of nearby levels. Experimentally, we observe the same trend, with the 16-level behavior emerging at higher magnetic fields  (45 G), unlike in the transverse field case where it was seen at lower value of magentic field. These results could be leveraged for quantum state engineering, optical switching, or sensitive detection of anisotropy in atomic and molecular systems.\\
\section{Equations}\label{eq}
\begin{figure}[htp!]
	\centering
	\includegraphics[width=0.5\textwidth]{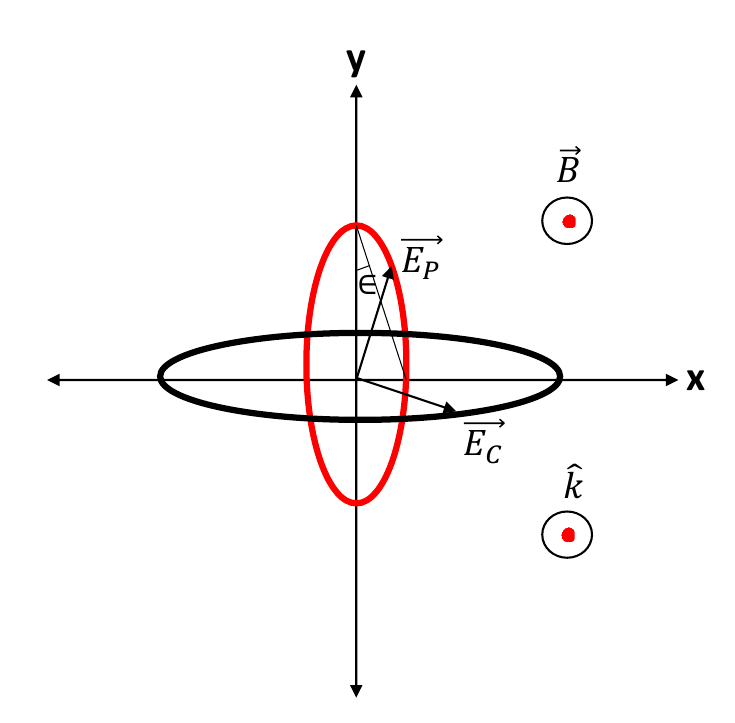}
	\caption{The ellipticity vectors of probe (red) and pump (black) along with the direction of magnetic field ($\vec{B}$) and light propagation vector ($\hat{\text{k}}$).  }
	\label{fig2}
\end{figure}

We consider the light fields to be planes waves with elliptical polarisation,
\begin{eqnarray}
	\vec{E}= \vec{E_p} (e^{i \omega_p t} + e^{- i \omega_p t})+\vec{E_c} (e^{i \omega_c t}+e^{-i \omega_c t})
	\label{eq1}
\end{eqnarray}
where, $\omega_p$ and $\omega_c$ are the frequency of the probe and pump respectively. The spatial dependence of the phase ($e^{\pm i k z}$) part is neglected using the Dipole approximation. The vector amplitudes which includes polarisation are,
\begin{eqnarray}
	\vec{E_p}= E_p \hat{e}_p ~~~~~~ \vec{E_c}= E_c \hat{e}_c
	\label{eq2}
\end{eqnarray}
Where, $\hat{e}_p$ and $\hat{e}_c$ are represented in circular basis,
\begin{eqnarray}
	\hat{e}_p = - \cos(\epsilon-\frac{\pi}{4})~\hat{e}_+ + \sin(\epsilon-\frac{\pi}{4})~\hat{e}_-
	\label{eq3}
\end{eqnarray}
\begin{eqnarray}
	\hat{e}_c =  - \sin(\epsilon-\frac{\pi}{4})~\hat{e}_+ - \cos(\epsilon-\frac{\pi}{4})~\hat{e}_-
	\label{eq4}
\end{eqnarray}
Where, $\epsilon$ is the ellipticity and $\hat{e}_{\pm}=\mp(\frac{\hat{e}_x \pm i\hat{e}_y}{\sqrt{2}})$ are the circular basis. Note that the constraint $\hat{e}_p~.~\hat{e}_c=0$ allows us to define only one ellipticity parameter $\epsilon$ to represent the polarisation state of both the light fields (Fig.\ref{fig2}). Ellipticity is defined within the range $\frac{-\pi}{4} \leq \epsilon \leq \frac{\pi}{4}$, for which $\epsilon=0$ represents linearly polarised states and $\epsilon=\pm\frac{\pi}{4}$ represents circular polarisations. The fields in the new basis are,
\begin{eqnarray}
	\vec{E_p}= E^+_p~\hat{e}_++E^-_p~\hat{e}_-, ~~ \vec{E_c}= E^+_c~ \hat{e}_++E^-_c~\hat{e}_-
	\label{eq5}
\end{eqnarray}
where, $E^+_p=E_p \sin(\epsilon-\frac{\pi}{4})$ and so on.
The total Hamiltonian for the system is defined as,
\begin{eqnarray}
	H=H_0+H_I+H_B
	\label{eq6}
\end{eqnarray}
$H_0$ and $H_I$ represents bare state and interaction Hamiltonian respectively. $H_B$ represents the energy corrections due to the magnetic field. For a 16 level system using the rotating wave approximation (RWA) we have,

\begin{eqnarray}
	H_0= \hbar (\delta_p-\delta_c) \sum_{i=4}^{8} \ket{i}\bra{i}+
	\hbar(\delta_p-\Delta
	\label{eq7}
	\\\nonumber
	 -k\text{v}) \sum_{i=9}^{11} \ket{i}\bra{i}	+\hbar (\delta_p-k\text{v}) \sum_{i=12}^{16} \ket{i}\bra{i}
\end{eqnarray} 
Where, $\delta_p$ and $\delta_c$ are the frequency detuning of the probe and pump from the transitions $\ket{F=1}\rightarrow\ket{F'=2}$ and $\ket{F=2}\rightarrow\ket{F'=2}$ respectively. $\Delta$ is the frequency gap between the excited hyperfine states($\ket{F'=2}$ and $\ket{F'=1}$).
\begin{eqnarray}
	\nonumber
	H_I=-\frac{\hbar}{2} 
	[\Omega^+_p(c_{1,10}\ket{1}\bra{10}
	+c_{1,14}\ket{1}\bra{14}
	\\
	\nonumber
	+c_{2,11}\ket{2}\bra{11}
	+c_{2,15}\ket{2}\bra{15}
	+c_{3,16}\ket{3}\bra{16})
	\\
	\nonumber
	+\Omega^-_p(c_{1,12}\ket{1}\bra{12}
	+c_{2,9}\ket{2}\bra{9}
	+c_{2,13}\ket{2}\bra{13}
		\\
	\nonumber
	+c_{3,10}\ket{3}\bra{10}
	+c_{3,14}\ket{3}\bra{14})
	+\Omega^+_c(c_{4,9}\ket{4}\bra{9}
		\\
	\nonumber
	+c_{4,13}\ket{4}\bra{13}
	+c_{5,10}\ket{5}\bra{10}
	+c_{5,14}\ket{5}\bra{14}
		\\
	\nonumber
	+c_{6,15}\ket{6}\bra{15}
	+c_{6,11}\ket{6}\bra{11}
	+c_{7,16}\ket{7}\bra{16})
		\\
	\nonumber
	+\Omega^-_c(c_{5,12}\ket{5}\bra{12}
	+c_{6,9}\ket{6}\bra{9}
	+c_{6,13}\ket{6}\bra{13}
		\\
	\nonumber
	+c_{7,10}\ket{7}\bra{10}
	+c_{7,14}\ket{7}\bra{14}
	+c_{8,11}\ket{8}\bra{11}\\
	\nonumber
	+c_{8,15}\ket{7}\bra{16})]+ h.c.
\end{eqnarray} 
$\Omega^{\pm}_p$ and $\Omega^{\pm}_c$ are the reduced Rabi frequencies defined as,
\begin{eqnarray}
	\Omega^{\pm}_p=\frac{E^{\pm}_p \bra{J}|e \textbf{r}| \ket{J'}}{\hbar}~~~\Omega^{\pm}_c=\frac{E^{\pm}_c \bra{J}|e \textbf{r}| \ket{J'}}{\hbar}~~~
	\label{eq8}
\end{eqnarray}
$c_{i,j}$ is the Clebsch-Gordan coefficient which defines the transition strengths between the Zeeman levels \cite{steck2001rubidium}.The Hamiltonian for the magnetic part is given by, 

\begin{eqnarray}
	H_B=\mu_B B \sum_{i=1}^{16} m_{F_i} g_{F_i} \ket{i}\bra{i}
	\label{eq9}
\end{eqnarray}

Where, $m_{F_i}$ and $g_{F_i}$ are the magnetic quantum number and Lande' g factor associated with the state $\ket{i}$ respectively. $\mu_B$ is the Bohr magneton and B is the strength of the magnetic field. Similarly, the Hamiltonian for a 13 level system can also be written. 
Time evolution of the system within Density matrix formulation is given by,
\begin{eqnarray}
	\nonumber
	\dot{\rho}(\delta_c,\text{v}) = -\frac{i}{\hbar} [H(\delta_c,\text{v}),\rho(\delta_c,\text{v})]-\frac{1}{2}\{\Gamma ,\rho(\delta_c,\text{v})\}\\
	-\gamma\rho(\delta_c,\text{v})+ R
	\label{eq10}
\end{eqnarray}
Where, $\Gamma$ (of the order of MHz) is the relaxation matrix which incorporates the spontaneous decays out of the excited state. $\gamma$ is the decay rate due to collisions of atoms which is typically of the order of few KHz at room temperature and R is the re-population matrix.  The velocity distribution function of the atoms as a function of temperature is given by,
\begin{eqnarray}
	f(\text{v})=\sqrt{\frac{m}{2\pi K_B T}} exp(-\frac{mv^2}{2 K_B T})
	\label{eq11}
\end{eqnarray}

First, the population distribution among the ground states for the 13- and 16-level systems is studied by solving equation \ref{eq10}. Secondly, in order to obtain the density matrix elements for each specific velocity in the range of -300 m/s to 300 m/s, the steady state solution is obtained ($\dot{\rho}=0$). To obtain the final findings, these density matrix elements are then averaged using f(v) as a kernel.
\begin{eqnarray}
	\bar{\rho}_{i,j}(\delta_c)= \int d\text{v} ~f(\text{v})~~ \rho(\delta_c,\text{v})
	\label{eq12}
\end{eqnarray}

\section{Results and Discussion}\label{re}
\begin{figure}[htp]
	\centering
	\includegraphics[width=0.45\textwidth]{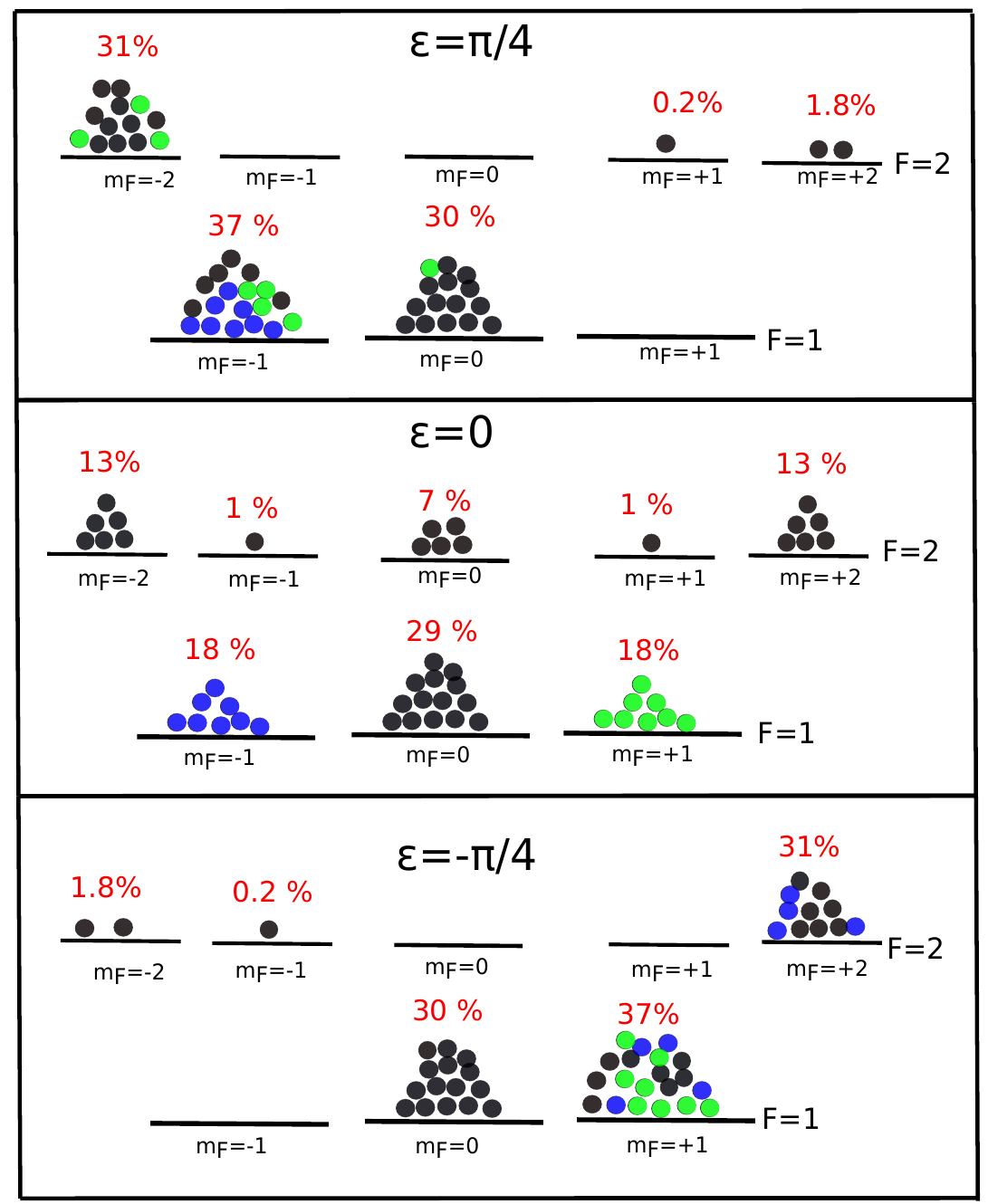}
	\caption{Schematic showing the population distrbuted at the Zeeman sub-levels of $F=1$ and  $F=2$ ground states for 13-level model. $\epsilon=0$ is the case for linearly polarized (orthogonal) pump and probe. $\epsilon=\pm\frac{\pi}{4}$ means pure circular polarizations. }
	\label{pop13}
\end{figure}

\begin{figure}[htp]
	\centering
	\includegraphics[width=0.5\textwidth]{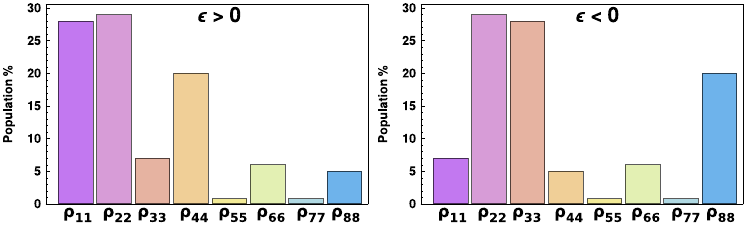}
	\caption{ Gound state populations at the Zeeman sub-levels of $F=1$ and  $F=2$ ground states for 13-level model for postive and negative ellipticity.  }
	\label{bar13}
\end{figure}

\subsection{13 level system}

\subsubsection{Ground state population}

 In this section, we investigate the population that are trapped in the Zeeman manifold of ground states in absence of magnetic field for the 13-level model.
Figures \ref{pop13} and \ref{bar13} show the normalized populations for different polarizations of the light fields at various values of $\epsilon$. By solving the time-dependent equation, we obtain the populations that remain trapped in the ground state once the system has reached steady state ($\tau_{steady}\approx 10 \Gamma^{-1}$). Throughout the evaluations, the probe and pump intensity are kept fixed at 0.3 mW/cm$^2$ and 161.2 mW/cm$^2$ respectively.\\
For $\epsilon = 0$ (see Fig.~\ref{pop13}), both the pump and probe fields are linearly polarized (orthogonal to each other) and provide equal power to the $\sigma^{\pm}$ transitions. Under these conditions, the system exhibits alignment in both ground states ($F=1$ and $F=2$). A portion of the population becomes trapped in the Zeeman sub-levels of $F=2$, which is attributed to coherent population trapping (CPT) induced by the pump field. In this scenario, a generalized dark state can be defined as
\begin{eqnarray}
	\ket{D}= \sum_{i=1}^{8} a_i(\epsilon) \ket{i}\\
\end{eqnarray} 
A generalized dark state refers to a coherent superposition of the ground states $\ket{1}$, $\ket{2}$, $\ldots$, $\ket{8}$, with populations $\rho_{ii} = |a_i|^2$ for $i = 1, \ldots, 8$, as presented in Figures~\ref{pop13} and~\ref{bar13}. The probability amplitudes $a_i$ can be controlled by adjusting the parameter $\epsilon$ and depend on the coupling coefficients $c_{i,j}$ and the Rabi frequencies $\Omega_p^{\pm}$ and $\Omega_c^{\pm}$.\\

As the system is taken to the case of pure circular polarizations ($\epsilon = \frac{\pi}{4}$), where the pump is $\sigma^-$ and the probe is $\sigma^+$, the system shifts from alignment to orientation. Both ground states exhibit orientation; however, there is a key difference. The population trapped in the state $\ket{F=2, m_F=-2}$  ($\rho_{44}$)  arises because this state is not coupled to any light field for the given polarizations of the pump and probe. In contrast, the population trapped in the ground state $F=1$ is attributed to the formation of two dark states arisng from two $\Lambda$ systems(see Table\ref{tab1} entries 4 and 8).\\
For the opposite case ($\epsilon = -\frac{\pi}{4}$), where the pump is $\sigma^+$ and the probe is $\sigma^-$, similar population distribution is observed. In this configuration, the orientation is now towards the states $\ket{F=2, m_F=+2}$  ($\rho_{88}$)  and $\ket{F=1, m_F=+1}$  ($\rho_{33}$) . This highlights that the conversion from alignment to orientation  exhibits symmetry for these two polarization configurations and is governed by the formation of dark states.\\
The populations for the intermediate case are shown in Fig.~\ref{bar13}. For $\epsilon > 0$, that is, as the system transitions from alignment to orientation, the population shifts towards the states $\ket{F=2, m_F=-2}$ ($\rho_{44}$) and $\ket{F=1, m_F=-1}$ ($\ket{\rho_{11}}$) with a decrease in the population trapped in $\ket{F=1, m_F=+1}$ ($\rho_{33}$). Similarly, for $\epsilon < 0$, there is an increase in the populations of $\ket{F=2, m_F=+2}$ ($\rho_{88}$) and $\ket{F=1, m_F=+1}$ ($\rho_{33}$), accompanied by a simultaneous decrease in the population of $\ket{F=1, m_F=+1}$ ($\rho_{11}$).\\ From this analysis, we predict that the populations trapped in the $F=1$ Zeeman manifold at two-photon resonance (a condition necessary for the formation of dark states) when $\epsilon = 0$, will convert to a bright state as the system transitions from alignment to orientation. In particular, for $\epsilon > 0$, the state $\ket{F=1, m_F=+1}$ ($\ket{3}$) becomes an absorbing (bright) state, whereas for $\epsilon < 0$, it is the state $\ket{F=1, m_F=-1}$ ($\ket{1}$) that becomes absorbing. To observe this conversion at intermediate values of $\epsilon$, the system must be subjected to an external magnetic field. The magnetic field lifts the degeneracy of the Zeeman sublevels, allowing the effects of the various $\Lambda$ systems to become apparent.\\
In the next section, we present the steady-state solution and analyze the probe transmission in the presence of a magnetic field at different values of  $\epsilon$.

\begin{figure}[htp]
	\centering
	\includegraphics[width=0.38\textwidth]{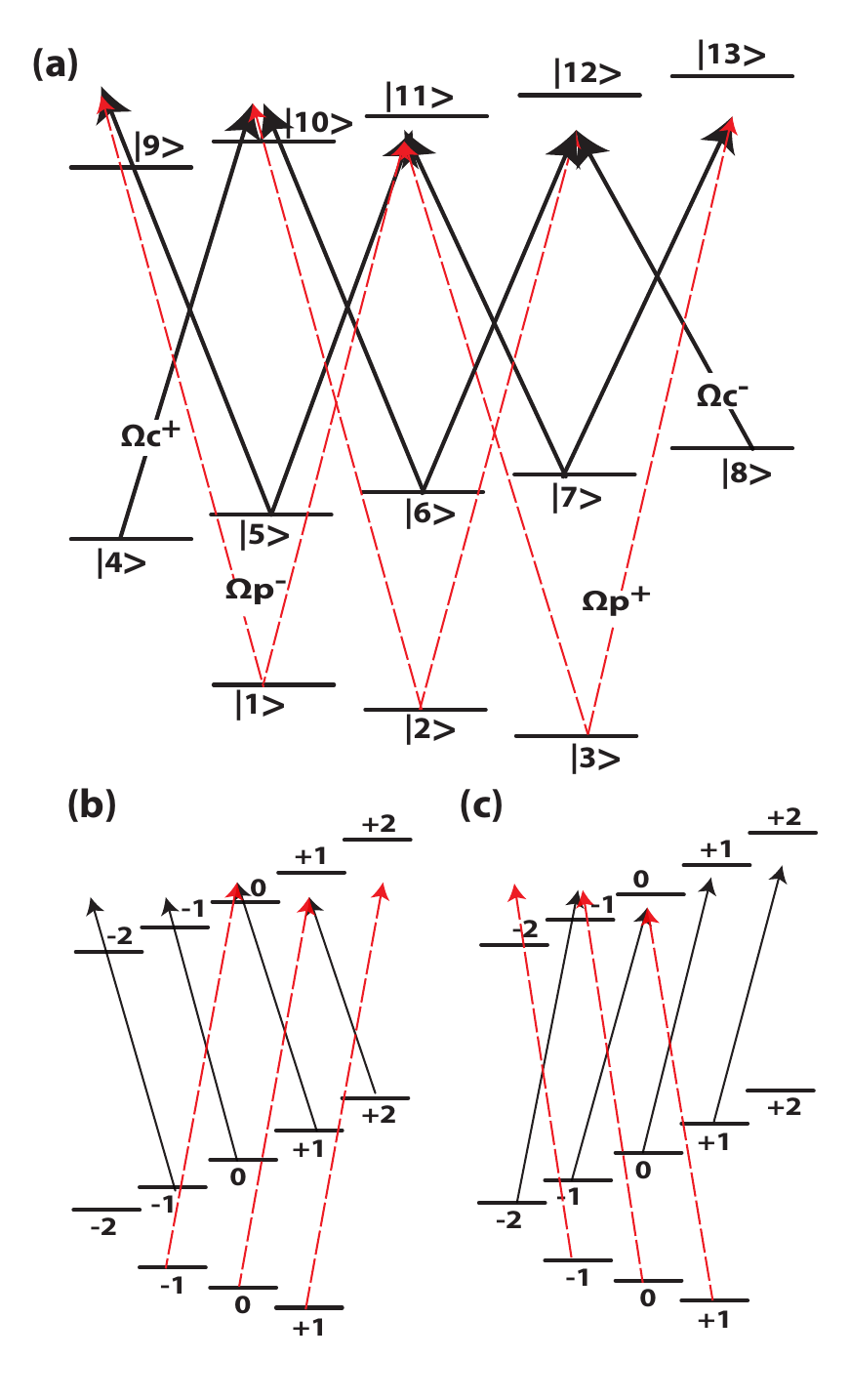}
	\caption{(a) The energy level diagram showing 13 levels. The red (dashed) line represents the weak probe and the black (solid) line is used to represent the strong pump. Energy level diagrams for 13-level system showing the formation of two $\Lambda$ systems each in case of (b). $\epsilon =\frac{\pi}{4}$ and (c). $\epsilon = -\frac{\pi}{4}$}
	\label{fig4}
\end{figure}

\begin{table*}[ht]
	\centering
	\caption{Total $\Lambda$ systems in 13-level system and their corresponding two photon resonance position.}

	\begin{tabular}{ccc}\hline
		Sl.No.&$\Lambda$ system&Position of two photon resonance ($\Delta_g=\frac{\mu_B B}{2\hbar}$)\\\hline
		1& $\ket{1}\rightarrow\ket{9}\leftarrow\ket{5}$ & -2$\Delta_g$\\ 
		2&$\ket{1}\rightarrow\ket{11}\leftarrow\ket{5}$ & -2$\Delta_g$\\ 
		3&$\ket{2}\rightarrow\ket{10}\leftarrow\ket{4}$ & -2$\Delta_g$\\ 
		4&$\ket{1}\rightarrow\ket{11}\leftarrow\ket{7}$ & 0\\
		5&$\ket{2}\rightarrow\ket{10}\leftarrow\ket{6}$ & 0\\
		6&$\ket{2}\rightarrow\ket{12}\leftarrow\ket{6}$ & 0\\
		7&$\ket{3}\rightarrow\ket{11}\leftarrow\ket{5}$ & 0\\
		8&$\ket{2}\rightarrow\ket{12}\leftarrow\ket{8}$ & +2$\Delta_g$\\
		9&$\ket{3}\rightarrow\ket{11}\leftarrow\ket{7}$ & +2$\Delta_g$\\
		10&$\ket{3}\rightarrow\ket{13}\leftarrow\ket{7}$ & +2$\Delta_g$\\  
		\hline

	\end{tabular}
	\label{tab1}
\end{table*} 

\begin{figure*}
	\centering
	\includegraphics[width=\textwidth]{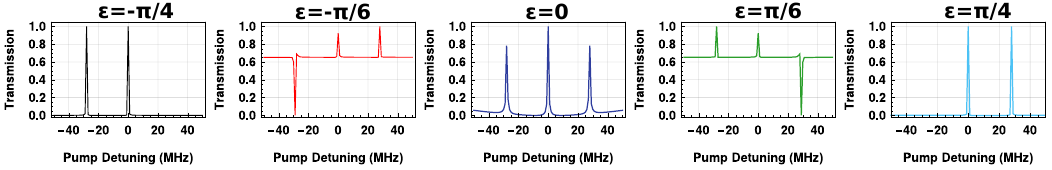}
	\caption{Numerical results showing transmission of the weak probe as a function of pump detuning ($\delta_c$) at fixed value of B=20G, probe intensity = 0.3 mW/cm$^2$ and pump intensity = 161.2 mW/cm$^2$. Three transmission peaks for ellipticity ($\epsilon$)=$0$. Two transmission and one absorption (at blue detuned part) for $\epsilon=+30^0$~($\frac{\pi}{6}$). Two transmission peaks for $\epsilon=+45^0$~($\frac{\pi}{4}$). Two transmission and one absorption (at red detuned part) for $\epsilon=-30^0$($-\frac{\pi}{6}$). Two transmission peaks for $\epsilon=-45^0 $($-\frac{\pi}{4}$).}
	\label{fig3}
\end{figure*}

\subsubsection{Steady state solution}

For the 13-level system, the response of the probe field by introducing magnetic perturbation is found by calculating the susceptibility given by,

\begin{eqnarray}
	\nonumber
	\chi_p(\delta_c)=\frac{2 N |d|^2 }{\hbar \epsilon_0}[(\Omega^+_p)^{-1}(c_{1,11}~\bar{\rho}_{1,11}+c_{2,12}~\bar{\rho}_{2,12}\\
	\nonumber
	+c_{3,13}~\bar{\rho}_{3,13})+(\Omega^-_p)^{-1}(c_{1,9}~\bar{\rho}_{1,9}
\\+c_{2,10}~\bar{\rho}_{2,10}+c_{3,11}~\bar{\rho}_{3,11})]
\label{eq13}
\end{eqnarray}
Where, N is the number density and $\epsilon_0$ is the permittivity of free space and $d=\bra{J}|e \textbf{r}| \ket{J'}$. Finally, the transmission of the probe is calculated using the imaginary part of the susceptibility,
\begin{eqnarray}
	\text{T}=\text{T}_0 e^{-k |\text{Im}(\chi_p)|~\text{L}}
	\label{eq14}
\end{eqnarray}

Where, L is the length of the cell containing the atomic vapor.
The numerical transmission spectra are shown in Fig.~\ref{fig3}. Calculations were performed for five different values of ellipticity ($\epsilon = 0, \pm \pi/6, \pm \pi/4$) with a fixed magnetic field of B$ = 20$ G and the probe and pump intensities were kept fixed at 0.3 mW/cm$^2$ and 161.2 mW/cm$^2$, respectively.\\
In Fig.~\ref{fig3}, the spectra for $\epsilon = 0$—where both the $\sigma^{\pm}$ components of the pump and probe have equal intensity—show three distinct transmission peaks. Each of these peaks arises from the superposition of different two-photon resonances associated with various $\Lambda$ systems (see Fig.~\ref{fig4} (a)). In total, there are ten $\Lambda$ systems (see Table~\ref{tab1}) present in this configuration, with their two-photon resonance conditions satisfied at $\delta_c = 0$ and $\delta_c = \pm 2\Delta_g$, where $\Delta_g = \frac{\mu_B B}{2\hbar}$. Of these ten $\Lambda$ systems, four contribute to the transmission at $\delta_c = 0$, while each of the transmission peaks at $\delta_c = \pm 2\Delta_g$ results from the superposition of three such $\Lambda$ systems.\\

The result for the case of $\epsilon=\frac{\pi}{6}$ show that the two photon resonance at $\delta_c=2\Delta_g$ i.e, the blue detuned peak flips and converts to an absorption. This suggests that the population in the state $\ket{F=1, m_F=+1}$ ($\ket{3}$) is undergoing absorption near the two-photon resonance condition with a small light shift, which is of the order of $(\Omega^+_c)^2/\delta_p$. There are two $\Lambda$ systems associated with $\ket{3}$ for which two-photon resonance condition is fulfilled at $\delta_c=2\Delta_g$ (see Table~\ref{tab1}). For these two $\Lambda$ systems, $\ket{3} \rightarrow \ket{11} \leftarrow \ket{7}$ and $\ket{3} \rightarrow \ket{13} \leftarrow \ket{7}$, which both satisfy the two-photon resonance condition at $\delta_c = 2\Delta_g$, we can define the following two sets of dark and bright states:
\begin{eqnarray}
	\ket{D1} = \Omega_c^+ \ket{3} - \Omega_p^+ \ket{7},~
	\ket{B1} = \Omega_p^+ \ket{3} + \Omega_c^+ \ket{7}\\
	\ket{D2} = \Omega_c^- \ket{3} - \Omega_p^- \ket{7},~
	\ket{B2} = \Omega_p^- \ket{3} + \Omega_c^- \ket{7}
\end{eqnarray}
where $(\Omega_c^+)^2 + (\Omega_p^+)^2 = 1$ and $(\Omega_c^-)^2 + (\Omega_p^-)^2 = 1$, ensuring normalization.With these sets of dark and bright states, the overlap between the dark state of one $\Lambda$ system and the bright state of the other can be quantified by the product $I_D \equiv$$|\langle D_1 | B_2 \rangle|^2$ (or equivalently $|\langle D_2 | B_1 \rangle|^2$) [reference].\\ 

For this case, the expression is,
\begin{eqnarray}
I_D = (\Omega_p^+~\Omega_c^- - \Omega_p^-~\Omega_c^+)^2
\end{eqnarray}
For $\epsilon=0$, since $\Omega_p^+ = \Omega_p^-$ and $\Omega_c^+ = \Omega_c^-$, it follows that $I_D = 0$. This indicates the existence of a perfect dark state for $\epsilon=0$, meaning the system is in a coherent superposition that is completely decoupled from the light fields leading to transmission at two photon resonance. However, as $\epsilon$ is varied towards positive values, $I_D$ begins to increase because the product $\Omega_p^+\Omega_c^-$ increases while $\Omega_p^-\Omega_c^+$ simultaneously decreases. This indicates that the system becomes increasingly likely to convert from transmission to absorption due to the enhanced overlap between the dark and bright states.\\
For the case of pure circular polarization, $\epsilon = \frac{\pi}{4}$ (with the pump being $\sigma^-$ and the probe being $\sigma^+$), the energy level diagram shown in Fig.~\ref{fig3}(b) highlights the presence of two $\Lambda$ systems. These two $\Lambda$ systems correspond to the two transmission peaks observed at $\delta_c = 0$ and $\delta_c = 2\Delta_g$ in the results (see Fig.~\ref{fig4}).\\
For the case of $\epsilon = -\frac{\pi}{6}$, symmetrical results are observed, with the conversion from transmission to absorption occurring at $\delta_c = -2\Delta_g$. This corresponds to absorption by the population in the $\ket{F=1, m_F=-1}$ ($\ket{1}$) state at two-photon resonance. For $\epsilon = -\frac{\pi}{4}$, two transmission peaks are observed at $\delta_c = 0$ and $\delta_c = -2\Delta_g$, which result from the two $\Lambda$ systems present in this configuration, as illustrated in Fig.~\ref{fig3}(c). In the earlier discussion, the alignment-to-orientation conversion was found to be symmetric for both $\epsilon > 0$ and $\epsilon < 0$ in the degenerate case. This symmetry is also reflected in the steady-state solutions, even in the presence of a magnetic field.\\
To understand the effect of the nearby hyperfine level, particularly that of $F'=1$, we next present a similar analysis using the 16-level model in the next section.

\subsection{16 level system}

\subsubsection{Ground state population}
\begin{figure}
	\centering
	\includegraphics[width=0.45\textwidth]{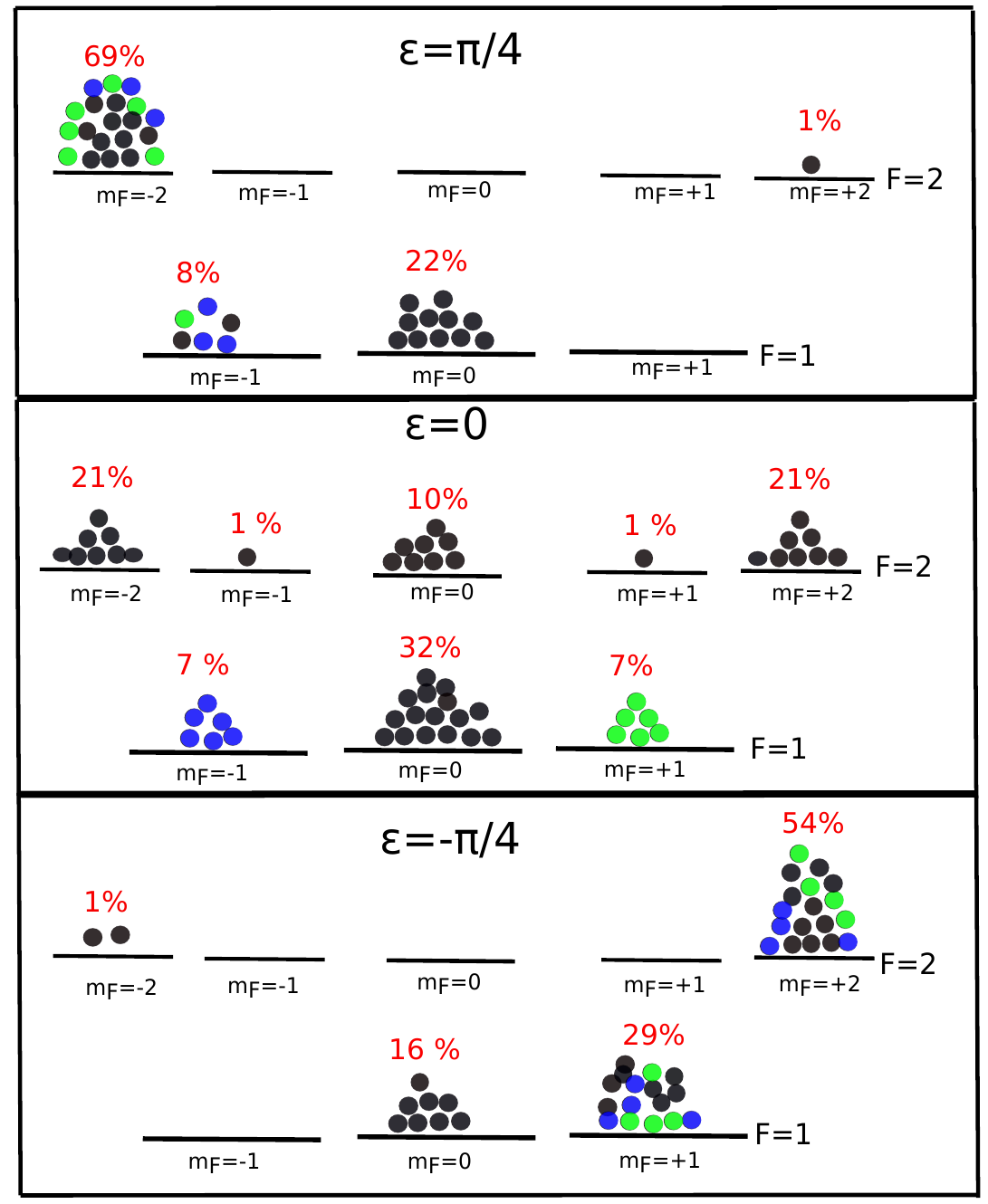}
	\caption{Schematic showing the population distrbuted at the Zeeman sub-levels of $F=1$ and  $F=2$ ground states for 16-level model. $\epsilon=0$ is the case for linearly polarized (orthogonal) pump and probe. $\epsilon=\pm\frac{\pi}{4}$ means pure circular polarizations.}
	\label{pop16}
\end{figure}
\begin{figure}
	\centering
	\includegraphics[width=0.5\textwidth]{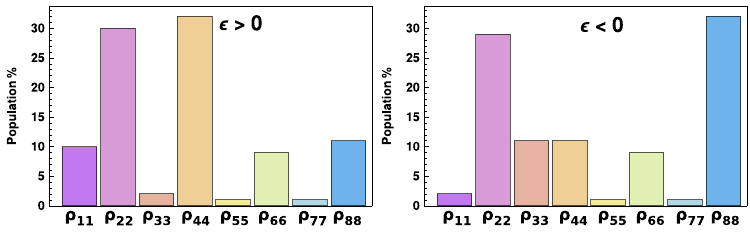}
	\caption{Gound state populations at the Zeeman sub-levels of $F=1$ and  $F=2$ ground states for 13-level model for postive and negative ellipticity. }
	\label{bar16}
\end{figure}

Similar to the case of the 13-level system, the inclusion of the Zeeman sub-levels of $F'=1$ for $\epsilon=0$ results in alignment of the populations in both ground states, $F=1$ and $F=2$ as shown in Fig.~\ref{pop16}. However, there is a notable difference in the degree of alignment: a larger fraction of the population is now trapped in the $F=2$ manifold, whereas in the 13-level case, the majority was trapped in $F=1$. This reversal suggests a stronger CPT effect induced by the pump beam, owing to the involvement of the additional sublevels associated with $F'=1$.\\
As the system approaches the pure circular polarization case, $\epsilon = \pm \frac{\pi}{4}$, the conversion from alignment to orientation becomes evident. However, there is a pronounced asymmetry with respect to the sign of the ellipticity. Specifically, a higher fraction of the population is trapped in the $\ket{F=2, m_F=-2}$ state ($\rho_{44}$) for $\epsilon = +\frac{\pi}{4}$, compared to the population trapped in $\ket{F=2, m_F=+2}$ ($\rho_{88}$) for $\epsilon = -\frac{\pi}{4}$. This asymmetry highlights the influence of the Zeeman sub-levels of the new hyperfine state.\\
For intermediate values of $\epsilon$, the population distributions are shown in Fig.~\ref{bar16}. A decrease in population is observed in state $\ket{3}$ for $\epsilon > 0$ and in state $\ket{4}$ for $\epsilon < 0$. These results indicate that the probe beam will exhibit absorption in both cases; however, due to the asymmetric population distribution, the degree of absorption may differ. The behavior of this system in the presence of a magnetic field is investigated, and the results are discussed in the next section.

\begin{figure}[htp!]
	\centering
	\includegraphics[width=0.38\textwidth]{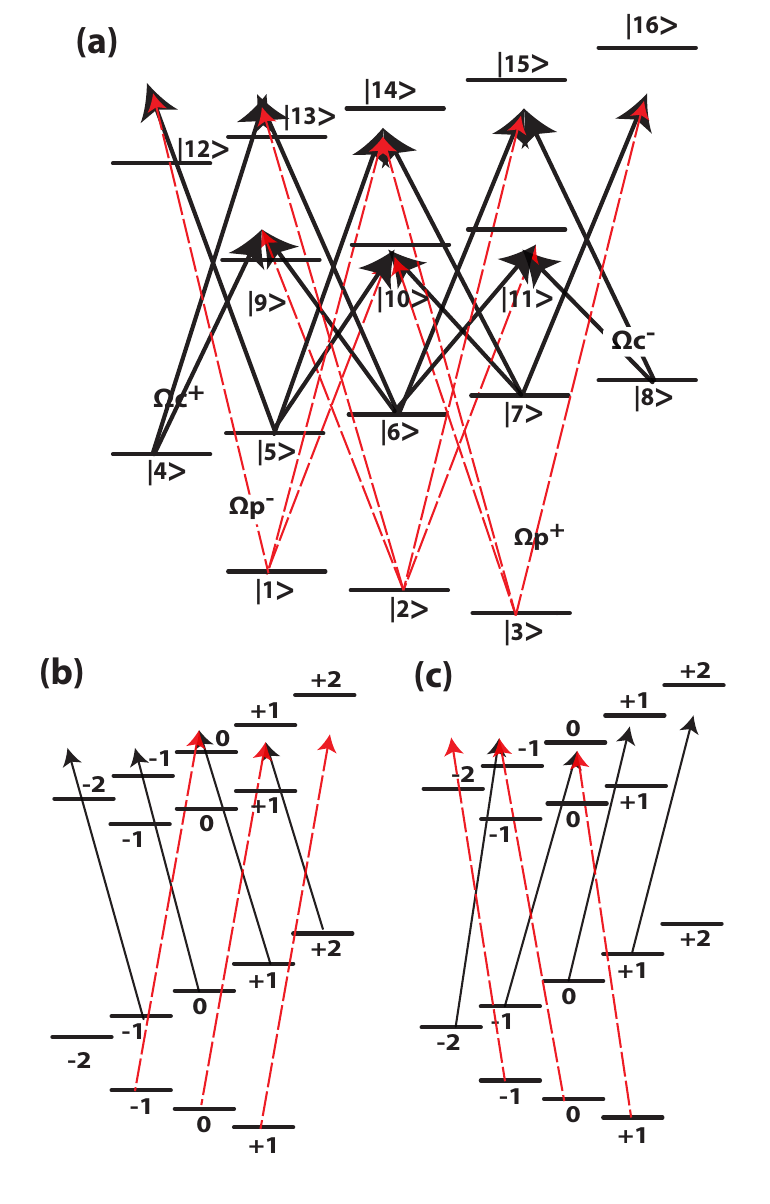}
	\caption{(a) The energy level diagram showing 16 levels. The red (dashed) line represents the weak probe and the black (solid) line is used to represent the strong pump. Energy level diagrams for 16-level systems showing $\Lambda$ systems each in case of (b). $\epsilon=$+45$^0$ and (c). $\epsilon$=-45$^0$}
	\label{fig6}
\end{figure}

\begin{table*}[htp]
	\centering
	\caption{Total $\Lambda$ systems in 16-level system and their corresponding two photon resonance position.}

	\begin{tabular}{ccc}\hline
		Sl.No.&$\Lambda$ system&Position of two photon resonance ($\Delta_g=\frac{\mu_B B}{2\hbar}$)\\\hline
		1& $\ket{1}\rightarrow\ket{12}\leftarrow\ket{5}$ & -2$\Delta_g$\\ 
		2&$\ket{1}\rightarrow\ket{14}\leftarrow\ket{5}$ & -2$\Delta_g$\\ 
		3&$\ket{1}\rightarrow\ket{10}\leftarrow\ket{5}$ & -2$\Delta_g$\\ 
		4&$\ket{2}\rightarrow\ket{13}\leftarrow\ket{4}$ & -2$\Delta_g$\\
		5&$\ket{2}\rightarrow\ket{9}\leftarrow\ket{4}$ & -2$\Delta_g$\\

		6&$\ket{1}\rightarrow\ket{14}\leftarrow\ket{7}$ & 0\\
		7&$\ket{1}\rightarrow\ket{10}\leftarrow\ket{7}$ & 0\\
		8&$\ket{2}\rightarrow\ket{13}\leftarrow\ket{6}$ & 0\\
		9&$\ket{2}\rightarrow\ket{9}\leftarrow\ket{6}$ & 0\\
		10&$\ket{2}\rightarrow\ket{15}\leftarrow\ket{6}$ & 0\\ 
		11&$\ket{2}\rightarrow\ket{11}\leftarrow\ket{6}$ & 0\\ 
		12&$\ket{3}\rightarrow\ket{14}\leftarrow\ket{5}$ & 0\\ 
		13&$\ket{3}\rightarrow\ket{10}\leftarrow\ket{5}$ & 0\\

		14&$\ket{2}\rightarrow\ket{15}\leftarrow\ket{8}$ & +2$\Delta_g$\\ 
		15&$\ket{}\rightarrow\ket{11}\leftarrow\ket{7}$ & +2$\Delta_g$\\ 
		16&$\ket{3}\rightarrow\ket{11}\leftarrow\ket{7}$ & +2$\Delta_g$\\ 
		17&$\ket{3}\rightarrow\ket{10}\leftarrow\ket{7}$ & +2$\Delta_g$\\ 
		18&$\ket{3}\rightarrow\ket{13}\leftarrow\ket{7}$ & +2$\Delta_g$\\  
		\hline

	\end{tabular}
	\label{tab2}
\end{table*}

\begin{figure*}[htp!]
	\centering
	\includegraphics[width=\textwidth]{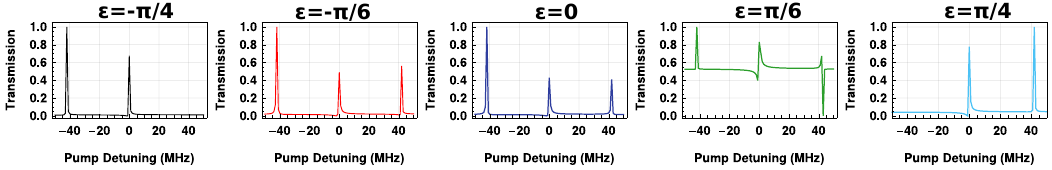}
	\caption{ Numerical results showing transmission of the weak probe as a function of pump detuning ($\delta_c$) at fixed value of B=30G, probe intensity = 0.3 mW/cm$^2$ and pump intensity = 161.2 mW/cm$^2$. Three transmission peaks for ellipticity ($\epsilon$) = 0. Two transmission and one absorption (at blue detuned part) for $\epsilon=+30^0$~($\frac{\pi}{6}$). Two transmission peaks for $\epsilon=+45^0$~($\frac{\pi}{4}$). Three transmission peaks for $\epsilon=-30^0$~($\frac{\pi}{6}$). Two transmission peaks for $\epsilon=-45^0$~(-$\frac{\pi}{4}$).}
	\label{fig5}
\end{figure*}

\subsubsection{Steady state solution}

The numerical calculations for the 16-level system are performed at a magnetic field strength of 30 G. The susceptibility of the probe field in this case is calculated as follows:
\begin{eqnarray}
\nonumber
\chi_p(\delta_c)=\frac{2 N |d|^2 }{\hbar \epsilon_0}[(\Omega^+_p)^{-1}(c_{1,10}~\bar{\rho}_{1,10}+\\
\nonumber
c_{1,14}~\bar{\rho}_{1,14}+c_{2,11}~\bar{\rho}_{2,11}+c_{2,15}~\bar{\rho}_{2,15}+\\ 
\nonumber
 c_{3,16}~\bar{\rho}_{3,16})+(\Omega^-_p)^{-1}(c_{1,12}~\bar{\rho}_{1,12}+c_{2,9}~\bar{\rho}_{2,9}+\\
c_{2,13}~\bar{\rho}_{2,13}+c_{3,10}~\bar{\rho}_{3,10}+c_{3,14}~\bar{\rho}_{3,14})]
\label{eq15}
\end{eqnarray}
The corresponding energy level diagram, including the Zeeman sub-levels, is shown in Fig.~\ref{fig6}(a).For this system, we find there are in total 18 $\Lambda$ systems (Table \ref{tab2}) out of which five each are responsible for transmission at $\delta_c=\pm 2 \Delta_g$ and the remaining eight superpose at $\delta_c=0$.
The numerical transmission spectra for this case are shown in Fig.~\ref{fig5}. For $\epsilon = 0$, corresponding to linearly polarized light fields, three distinct transmission peaks are observed at $\delta_c = 0$ and $\delta_c = \pm 2\Delta_g$.\\
As the ellipticity is tuned to $\epsilon = \frac{\pi}{6}$, similar to the behavior observed in the 13-level system, the transmission peak at $\delta_c = +2\Delta_g$ is converted to an absorption dip. This change indicates absorption by the population trapped in the state $\ket{F=1, m_F=+1}$ ($\rho_{33}$) as the system transitions from alignment to orientation. For this value of $\epsilon$, the peak at $\delta_c = 0$ also exhibits asymmetric line-shape features. This asymmetry can be attributed to the fact that the ground state $\ket{F=1, m_F=+1}$ also participates in the resonance at $\delta_c = 0$ through its involvement in the $\Lambda$ systems (see Table~\ref{tab2}).\\
For $\epsilon = \frac{\pi}{4}$, where the pump (probe) is $\sigma^-$ ($\sigma^+$) polarized, two transmission peaks are observed at $\delta_c = 0$ and $\delta_c = 2\Delta_g$. Figure~\ref{fig6}(b) shows the corresponding energy level diagram for this configuration, which, together with Table~\ref{tab2}, confirms the existence of these resonances. Similarly, for $\epsilon = -\frac{\pi}{4}$, where the pump (probe) is $\sigma^+$ ($\sigma^-$) polarized, two transmission peaks are observed at $\delta_c = 0$ and $\delta_c =- 2\Delta_g$. The relevant energy level diagram for this case is shown in Fig.~\ref{fig6}(c)\\
Unlike the 13-level system, for $\epsilon = -\frac{\pi}{6}$, the results do not show any conversion to absorption at $\delta_c = -2\Delta_g$. As the system transitions from alignment to orientation in this case, the symmetry observed in the 13-level system with respect to the sign of ellipticity is lost.\\
As discussed earlier, this asymmetry arises because the orientation in the 16-level system is inherently not symmetric. This observation can be understood as follows: As shown in Table~\ref{tab2}, the resonance at $\delta_c = -2\Delta_g$ results from the superposition of five $\Lambda$ systems. At finite detuning, for the $\sigma^+$ probe beam, the transitions $\ket{2} \rightarrow \ket{13}$ and $\ket{2} \rightarrow \ket{9}$ are both blue-detuned, which diminishes the influence of the $\ket{2} \rightarrow \ket{9}$ transition. In contrast, for the $\sigma^+$ probe polarization, the transition $\ket{2} \rightarrow \ket{15}$ is red-detuned while $\ket{2} \rightarrow \ket{11}$ is blue-detuned, leading to a stronger contribution from the state $\ket{11}$.\\
The $\Lambda$ systems involving $\ket{2} \rightarrow \ket{9} \leftarrow \ket{4}$ and $\ket{2} \rightarrow \ket{13} \leftarrow \ket{4}$ are responsible for the transmission peak at $\delta_c = -2\Delta_g$, and we therefore expect stronger transmission due to these pathways. On the other hand, the $\Lambda$ systems $\ket{2} \rightarrow \ket{11} \leftarrow \ket{8}$ and $\ket{2} \rightarrow \ket{15} \leftarrow \ket{8}$ contribute to the peak at $\delta_c = +2\Delta_g$, which is known to produce a more pronounced asymmetry in the resonance line shape~[ref]. Hence, the conversion from transmission to aborption is oberved at $\delta_c = +2\Delta_g$ and is absent for at $\delta_c = -2\Delta_g$. This detailed interplay between the detunings and the specific $\Lambda$ systems involved explains the loss of symmetry with respect to the sign of ellipticity in the 16-level system, in contrast to the behavior observed in the 13-level case.\\
In next section, we present an experiment performed in the $D_2$ line of $^{87}Rb$ at room temperature, where both the 13-level and 16-level results are observed at two different values of magnetic field. 
\section{Experiment on D$_2$ line of $^{87}$Rb}\label{exp}
\begin{figure}[htp]
	\centering
	\includegraphics[width=0.5\textwidth]{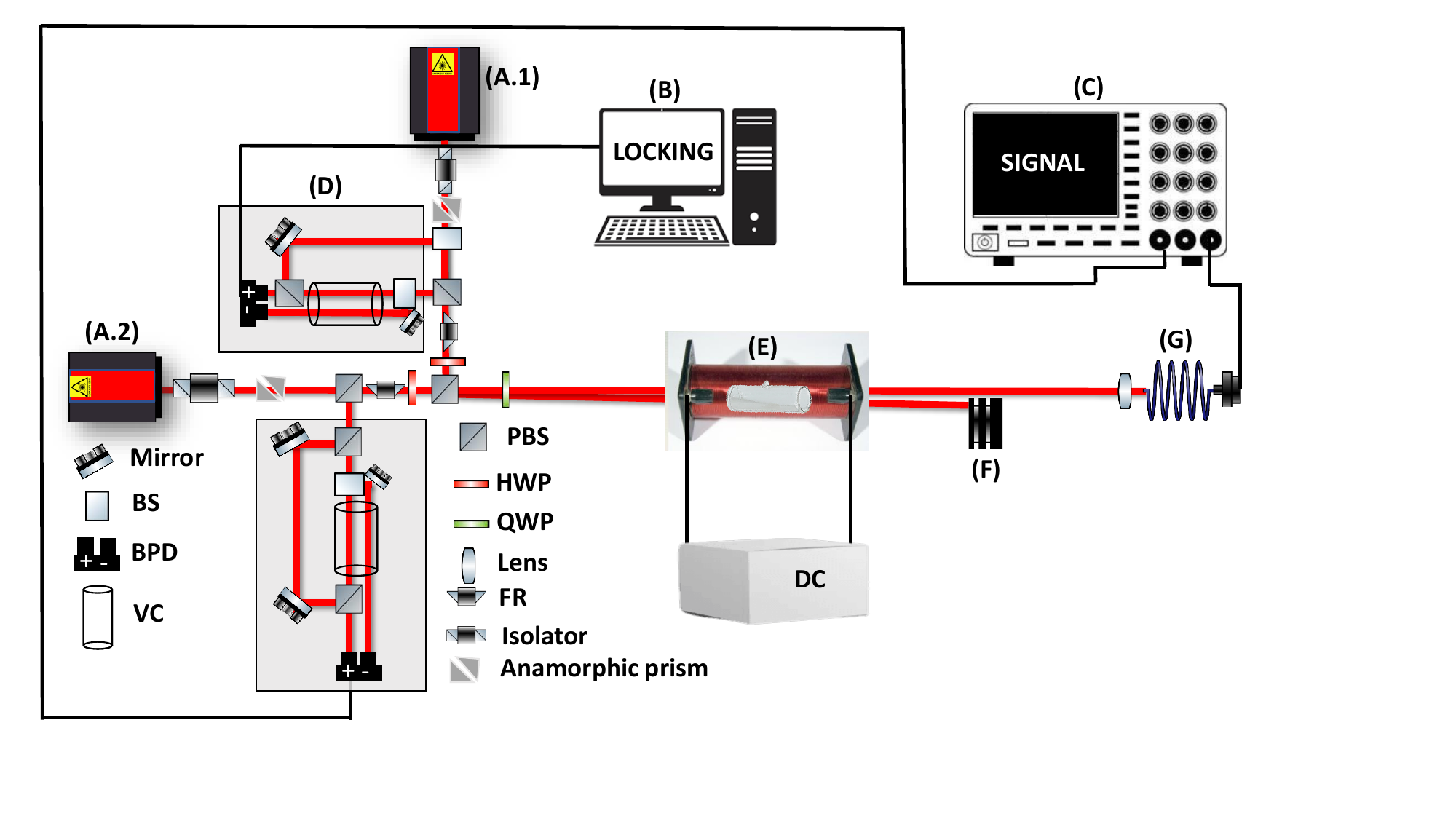}
	\caption{Experimental setup used to obtain the experimental results. (A.1 and A.2) The laser sources. (B) Laser locking module. (C) Digital Oscilloscope. (D) Saturated absorption spectroscopy (SAS) setup. BS (Beam Splitter); BPD (Balanced Photo Detector); VC (Vapor Cell) (E) Solenoid with the vapor cell. The strength of the magnetic field is controlled using an external DC power supply. (F) Beam dump (G) Ultra fast photo detector.  }
	\label{fig8}
\end{figure} 
\subsection{Experimental setup}

The optical setup shown in Fig.\ref{fig8} was used for obtaining the experimental results. Both the laser sources (probe and pump) are generated from external feedback diode lasers with central wavelength of 780 nm having linewidth of 1 MHz. The experiments were performed in the D$_2$ line of $^{87}$Rb atoms. Using the laser locking electronics (Toptica digilock 110), the probe laser was locked from F=1 to F'=2 transition and the frequency of the pump laser was scanned across the hyperfine transitions. To avoid any back reflection optical isolators were used. The beam profile was also made nearly circular using anamorphic prism pair. The beam diameter of both the lasers are 0.2 cm. The signals produced with saturated absorption spectroscopy (SAS) was used as a frequency reference scale for both the lasers. Both the lasers were also passed through a faraday rotator for pre control of the polarisation before it is combined with a polarizing beam splitter (PBS). The power of both the lasers were controlled using a combination of half wave plate (HWP) and PBS. The ellipticity of both the laser is simultaneously tuned using a quarter wave plate (QWP). The experimental cell is a Rb vapor cell having dimension of 25 x 75 mm. It was maintained at room temperature with a temperature controller system. A solenoid, with a length of 10 cm and a diameter of 6 cm is used for generating the magnetic field. It consists of 590 turns and was subjected to an applied current between 0-1.5 A. The entire chamber enclosed with $\mu$ metal sheets to nullify the earth's magnetic field. The detected signal is measured using an fiber coupled ultra-fast photo detector and observed in a 5 channel oscilloscope. 
\begin{figure}[htp!]
	\centering
	\includegraphics[width=0.3\textwidth]{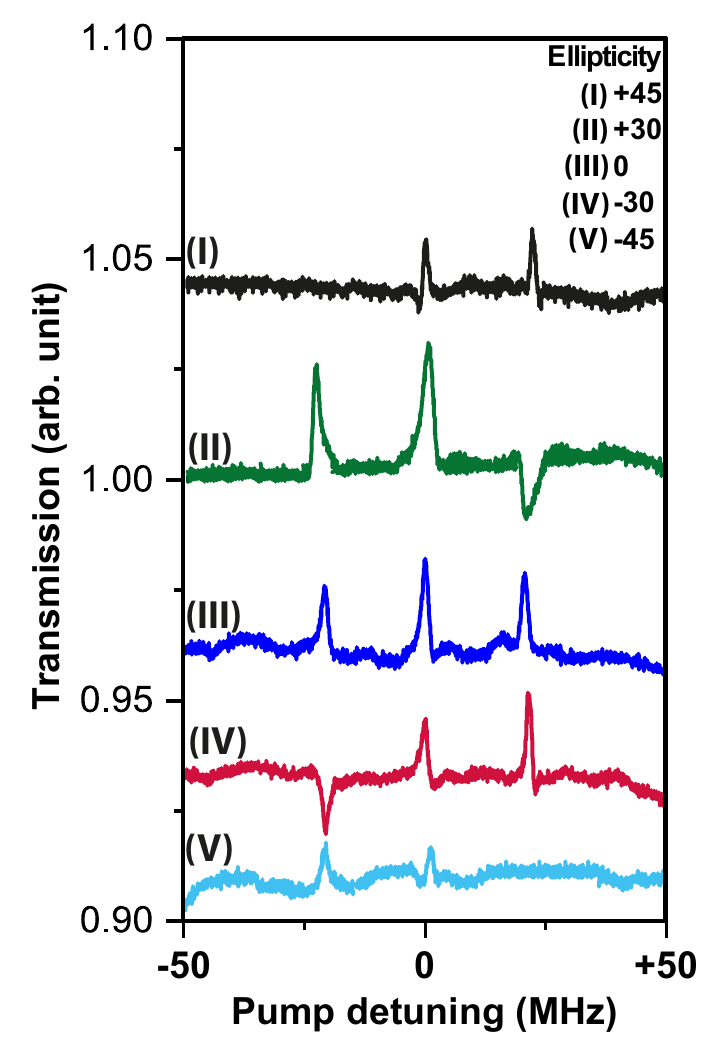}
	\caption{ Experimental results showing the transmission of the weak probe as a function of pump detuning at fixed value of B= 17G, probe intensity = 3.3 mW/cm$^2$ and pump intensity=161.2 mW/cm$^2$. Blue spectra (curve-III) is the result with  ellipticity ($\epsilon$)=0$^0$ in which three transmission peaks are observed. Green spectra and red spectra (curve-II and IV) are the results obtained for $\epsilon$=$\pm$30$^0$ where two transmission and one absorption peak is observed. Black and blue curve (curve-I and V) are the results obtained for $\epsilon$=$\pm$45$^0$ which shows two transmission peaks.}
	\label{fig9}
\end{figure} 

\subsection{Results}
In the experiment, the probe is locked at the $\ket{F=1}\rightarrow\ket{F'=2}$ transition, and the pump is scanned around $\ket{F=2}\rightarrow\ket{F'=2}$. The probe and pump intensity are kept fixed at 3.3 mW/cm$^2$ and 161.2 mW/cm$^2$ throughout the study.
\begin{figure}[htp]
	\centering
	\includegraphics[width=0.3\textwidth]{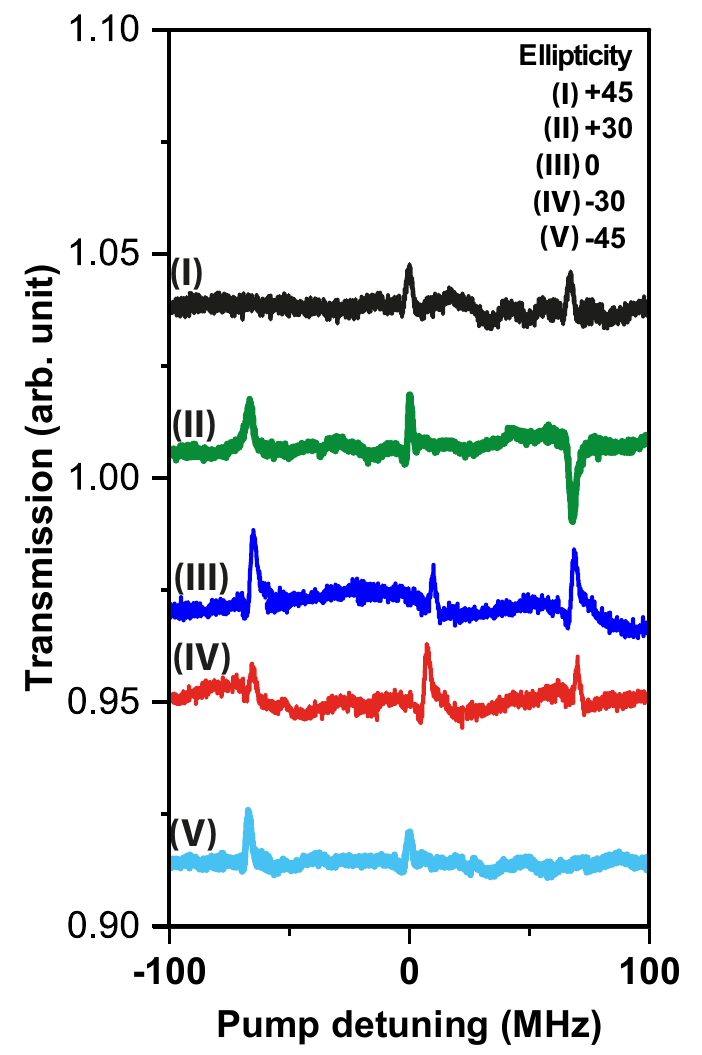}
	\caption{Experimental results showing the transmission of the weak probe as a function of pump detuning at fixed value of B= 45 G, probe intensity = 3.3 mW/cm$^2$ and pump intensity=161.2 mW/cm$^2$. Blue spectra (curve-III) is the result with  ellipticity ($\epsilon$)=0$^0$ in which  three transmission peaks are observed. Green spectra spectra (curve-II ) is the results obtained $\epsilon$=$+$30$^0$ which shows two transmission and one absorption peak. Red spectra (curve-IV ) is for $\epsilon$=$-$30$^0$, three transmission peaks are observed. Black and blue curve (curve-I and V) are the results obtained for $\epsilon$=$\pm$45$^0$ where two transmission peaks are observed.}
	\label{fig10}
\end{figure} 
The experimental results obtained for a magnetic field of 17 G are shown in Fig. \ref{fig9}. As discussed in the previous section, the ellipticity ($\epsilon$) of both beams is simultaneously varied using a quarter wave plate. For the case when $\epsilon$ = $0$, three transmission peaks are observed (curve-III). The peak positions of the three peaks are at $\delta_c=0,\pm 2\Delta_g$ ($\Delta_g=\frac{\mu_B B}{2\hbar}$).
 
As discussed earlier, altogether, 10 $\Lambda$ systems are responsible for these three transmission peaks. By tuning the ellipticity to + 30$^0$ ($\epsilon=\frac{\pi}{6}$), one of the peaks ($+2\Delta_g$) converts to absorption (curve-II). By further changing the ellipticity to + 45$^0$ (($\epsilon=\frac{\pi}{4}$)), only two transmission peaks are observed (curve-I). For - 30$^0$ ($\epsilon=-\frac{\pi}{6}$), the observation is similar to what is observed for + 30$^0$. However, the position of the absorption peak is observed on the red detuned part of the spectrum (curve-IV). Observations similar to + 45$^0$ are made when the ellipticity is tuned to - 45$^0$ (curve-V). In this case too, the only difference is in the position of the two peaks. For ellipticity higher than 45$^0$, the features repeated themselves, maintaining the symmetry we had observed for positive and negative values of ellipticity. The results are in complete agreement with the numerical results of the 13-level system. Hence, we find that the anisotropy created in the population distribution as the system transitions from alignment to orientation leads to optical anisotropy, resulting in absorption of the probe beam at two-photon resonance. These results also indicate that, at this value of magnetic field, the effect of the Zeeman sublevels of $\ket{F'=1}$ is negligible for the $D_2$ line.\\
The experimental observations for a magnetic field of 45 G are shown in Fig.\ref{fig10}. The reason for increasing the magnetic field was to create a situation where the effects of other hyperfine levels became more prominent. For $\epsilon$=0, three transmission peaks are observed similar to what is observed for low magnetic field (curve-I). However, the amplitudes of these three peaks are different. For ellipticity equal to + 30$^0$, an absorption peak on the blue detuned part and two transmission peaks are observed, as shown in curve-II of Fig.\ref{fig10}. Two transmission peaks are visible for both the case of circularly polarised light fields, i.e., for + 45$^0$ and -45$^0$ as shown in Fig.\ref{fig10}  (curve-I and V respectively). The main difference in the observations for this case as compared to the low magnetic field is the absence of symmetry with respect to the sign of ellipticity. The results produced here for B=45 G are in agreement with the 16-level model system. Hence, we conclude that for our model system of D$_2$ line of $^{87}$ Rb at higher magnetic field, nearby Zeeman states of $F'=1$ actively participate in two-photon processes.\\

\section*{Conclusion}\label{conc}
We have experimentally and theoretically investigated anisotropy-induced changes in probe transmission at two-photon resonance on the D$_2$ line of $^{87}$ Rb with varying ellipticity of pump and probe fields. Numerical simulations using 13- and 16-level Zeeman-resolved models were validated against room-temperature pump-probe measurements in a longitudinal magnetic field. For 13-level system, which excludes $F'=1$ excited hyperfine state, ground state population analysis revealed a symmetric alignment-orientation conversion with respect to ellipticity. At $\epsilon=0$ (orthogonal linear polarizations), three transmission peaks were observed originating from ten contributing $\Lambda$ systems. As ellipticity increased to $\pm\pi/6$, one of the detuned EIT peak converted into absorption, with sign determined by the handedness of the polarization. With pure circular polarization ($\epsilon=\pm\pi/4$), only two transmission peak were observed. Experimental spectra at B=17 G reproduced these features exactly, confirming the predicted symmetry between positive and negative ellipticity. For the 16-level system, which includes Zeeman sublevels of $F'=1$, the population distribution became asymmetric with respect to ellipticity, favoring one stretched ground state over the other. Steady-state calculations at B=30 G showed that the conversion from transmission to absorption occurred only for positive ellipticity. This loss of symmetry arises from the altered composition and detuning of the contributing $\Lambda$ systems involving $F'=1$. Experimental results at B=45 G displayed the same asymmetry, confirming the model prediction that nearby prediction that nearby hyperfine states can qualitatively change the two-photon anisotropy. The findings provide a predictive basis for tailoring coherent optical effects in multi-level atomic vapors, with potential applications in optical switching, quantum state preparation and anisotropy besed sensing. 

\bibliography{reference}

\providecommand{\noopsort}[1]{}\providecommand{\singleletter}[1]{#1}%
\begin{thebibliography}{42}%
\makeatletter
\providecommand \@ifxundefined [1]{%
 \@ifx{#1\undefined}
}%
\providecommand \@ifnum [1]{%
 \ifnum #1\expandafter \@firstoftwo
 \else \expandafter \@secondoftwo
 \fi
}%
\providecommand \@ifx [1]{%
 \ifx #1\expandafter \@firstoftwo
 \else \expandafter \@secondoftwo
 \fi
}%
\providecommand \natexlab [1]{#1}%
\providecommand \enquote  [1]{``#1''}%
\providecommand \bibnamefont  [1]{#1}%
\providecommand \bibfnamefont [1]{#1}%
\providecommand \citenamefont [1]{#1}%
\providecommand \href@noop [0]{\@secondoftwo}%
\providecommand \href [0]{\begingroup \@sanitize@url \@href}%
\providecommand \@href[1]{\@@startlink{#1}\@@href}%
\providecommand \@@href[1]{\endgroup#1\@@endlink}%
\providecommand \@sanitize@url [0]{\catcode `\\12\catcode `\$12\catcode
  `\&12\catcode `\#12\catcode `\^12\catcode `\_12\catcode `\%12\relax}%
\providecommand \@@startlink[1]{}%
\providecommand \@@endlink[0]{}%
\providecommand \url  [0]{\begingroup\@sanitize@url \@url }%
\providecommand \@url [1]{\endgroup\@href {#1}{\urlprefix }}%
\providecommand \urlprefix  [0]{URL }%
\providecommand \Eprint [0]{\href }%
\providecommand \doibase [0]{https://doi.org/}%
\providecommand \selectlanguage [0]{\@gobble}%
\providecommand \bibinfo  [0]{\@secondoftwo}%
\providecommand \bibfield  [0]{\@secondoftwo}%
\providecommand \translation [1]{[#1]}%
\providecommand \BibitemOpen [0]{}%
\providecommand \bibitemStop [0]{}%
\providecommand \bibitemNoStop [0]{.\EOS\space}%
\providecommand \EOS [0]{\spacefactor3000\relax}%
\providecommand \BibitemShut  [1]{\csname bibitem#1\endcsname}%
\let\auto@bib@innerbib\@empty
\bibitem [{\citenamefont {Fleischhauer}\ \emph {et~al.}(2005)\citenamefont
  {Fleischhauer}, \citenamefont {Imamoglu},\ and\ \citenamefont
  {Marangos}}]{fleischhauer2005electromagnetically}%
  \BibitemOpen
  \bibfield  {author} {\bibinfo {author} {\bibfnamefont {M.}~\bibnamefont
  {Fleischhauer}}, \bibinfo {author} {\bibfnamefont {A.}~\bibnamefont
  {Imamoglu}},\ and\ \bibinfo {author} {\bibfnamefont {J.~P.}\ \bibnamefont
  {Marangos}},\ }\bibfield  {title} {\bibinfo {title} {Electromagnetically
  induced transparency: Optics in coherent media},\ }\href@noop {} {\bibfield
  {journal} {\bibinfo  {journal} {Reviews of modern physics}\ }\textbf
  {\bibinfo {volume} {77}},\ \bibinfo {pages} {633} (\bibinfo {year}
  {2005})}\BibitemShut {NoStop}%
\bibitem [{\citenamefont {Finkelstein}\ \emph {et~al.}(2023)\citenamefont
  {Finkelstein}, \citenamefont {Bali}, \citenamefont {Firstenberg},\ and\
  \citenamefont {Novikova}}]{finkelstein2023practical}%
  \BibitemOpen
  \bibfield  {author} {\bibinfo {author} {\bibfnamefont {R.}~\bibnamefont
  {Finkelstein}}, \bibinfo {author} {\bibfnamefont {S.}~\bibnamefont {Bali}},
  \bibinfo {author} {\bibfnamefont {O.}~\bibnamefont {Firstenberg}},\ and\
  \bibinfo {author} {\bibfnamefont {I.}~\bibnamefont {Novikova}},\ }\bibfield
  {title} {\bibinfo {title} {A practical guide to electromagnetically induced
  transparency in atomic vapor},\ }\href@noop {} {\bibfield  {journal}
  {\bibinfo  {journal} {New Journal of Physics}\ }\textbf {\bibinfo {volume}
  {25}},\ \bibinfo {pages} {035001} (\bibinfo {year} {2023})}\BibitemShut
  {NoStop}%
\bibitem [{\citenamefont {Lezama}\ \emph {et~al.}(1999)\citenamefont {Lezama},
  \citenamefont {Barreiro},\ and\ \citenamefont
  {Akulshin}}]{lezama1999electromagnetically}%
  \BibitemOpen
  \bibfield  {author} {\bibinfo {author} {\bibfnamefont {A.}~\bibnamefont
  {Lezama}}, \bibinfo {author} {\bibfnamefont {S.}~\bibnamefont {Barreiro}},\
  and\ \bibinfo {author} {\bibfnamefont {A.}~\bibnamefont {Akulshin}},\
  }\bibfield  {title} {\bibinfo {title} {Electromagnetically induced
  absorption},\ }\href@noop {} {\bibfield  {journal} {\bibinfo  {journal}
  {Physical Review A}\ }\textbf {\bibinfo {volume} {59}},\ \bibinfo {pages}
  {4732} (\bibinfo {year} {1999})}\BibitemShut {NoStop}%
\bibitem [{\citenamefont {Liao}\ \emph {et~al.}(2020)\citenamefont {Liao},
  \citenamefont {Tu}, \citenamefont {Yang}, \citenamefont {Chen}, \citenamefont
  {Liu}, \citenamefont {Liang}, \citenamefont {Zhang}, \citenamefont {Yan},\
  and\ \citenamefont {Zhu}}]{liao2020microwave}%
  \BibitemOpen
  \bibfield  {author} {\bibinfo {author} {\bibfnamefont {K.-Y.}\ \bibnamefont
  {Liao}}, \bibinfo {author} {\bibfnamefont {H.-T.}\ \bibnamefont {Tu}},
  \bibinfo {author} {\bibfnamefont {S.-Z.}\ \bibnamefont {Yang}}, \bibinfo
  {author} {\bibfnamefont {C.-J.}\ \bibnamefont {Chen}}, \bibinfo {author}
  {\bibfnamefont {X.-H.}\ \bibnamefont {Liu}}, \bibinfo {author} {\bibfnamefont
  {J.}~\bibnamefont {Liang}}, \bibinfo {author} {\bibfnamefont {X.-D.}\
  \bibnamefont {Zhang}}, \bibinfo {author} {\bibfnamefont {H.}~\bibnamefont
  {Yan}},\ and\ \bibinfo {author} {\bibfnamefont {S.-L.}\ \bibnamefont {Zhu}},\
  }\bibfield  {title} {\bibinfo {title} {Microwave electrometry via
  electromagnetically induced absorption in cold rydberg atoms},\ }\href@noop
  {} {\bibfield  {journal} {\bibinfo  {journal} {Physical Review A}\ }\textbf
  {\bibinfo {volume} {101}},\ \bibinfo {pages} {053432} (\bibinfo {year}
  {2020})}\BibitemShut {NoStop}%
\bibitem [{\citenamefont {Alzetta}\ \emph {et~al.}(1976)\citenamefont
  {Alzetta}, \citenamefont {Gozzini}, \citenamefont {Moi}, \citenamefont
  {Orriols} \emph {et~al.}}]{alzetta1976experimental}%
  \BibitemOpen
  \bibfield  {author} {\bibinfo {author} {\bibfnamefont {G.}~\bibnamefont
  {Alzetta}}, \bibinfo {author} {\bibfnamefont {A.}~\bibnamefont {Gozzini}},
  \bibinfo {author} {\bibfnamefont {L.}~\bibnamefont {Moi}}, \bibinfo {author}
  {\bibfnamefont {G.}~\bibnamefont {Orriols}}, \emph {et~al.},\ }\bibfield
  {title} {\bibinfo {title} {An experimental method for the observation of rf
  transitions and laser beat resonances in oriented na vapour},\ }\href@noop {}
  {\bibfield  {journal} {\bibinfo  {journal} {Nuovo Cimento B}\ }\textbf
  {\bibinfo {volume} {36}},\ \bibinfo {pages} {5} (\bibinfo {year}
  {1976})}\BibitemShut {NoStop}%
\bibitem [{\citenamefont {Wu}\ \emph {et~al.}(2025)\citenamefont {Wu},
  \citenamefont {Li}, \citenamefont {Gallagher}, \citenamefont {Lawrie},\ and\
  \citenamefont {Wang}}]{wu2025coherent}%
  \BibitemOpen
  \bibfield  {author} {\bibinfo {author} {\bibfnamefont {S.}~\bibnamefont
  {Wu}}, \bibinfo {author} {\bibfnamefont {X.}~\bibnamefont {Li}}, \bibinfo
  {author} {\bibfnamefont {I.}~\bibnamefont {Gallagher}}, \bibinfo {author}
  {\bibfnamefont {B.}~\bibnamefont {Lawrie}},\ and\ \bibinfo {author}
  {\bibfnamefont {H.}~\bibnamefont {Wang}},\ }\bibfield  {title} {\bibinfo
  {title} {Coherent population trapping and spin relaxation of a silicon
  vacancy center in diamond at millikelvin temperatures},\ }\href@noop {}
  {\bibfield  {journal} {\bibinfo  {journal} {Physical Review B}\ }\textbf
  {\bibinfo {volume} {111}},\ \bibinfo {pages} {035421} (\bibinfo {year}
  {2025})}\BibitemShut {NoStop}%
\bibitem [{\citenamefont {Walker}\ \emph {et~al.}(2012)\citenamefont {Walker},
  \citenamefont {Arnold},\ and\ \citenamefont
  {Franke-Arnold}}]{walker2012trans}%
  \BibitemOpen
  \bibfield  {author} {\bibinfo {author} {\bibfnamefont {G.}~\bibnamefont
  {Walker}}, \bibinfo {author} {\bibfnamefont {A.}~\bibnamefont {Arnold}},\
  and\ \bibinfo {author} {\bibfnamefont {S.}~\bibnamefont {Franke-Arnold}},\
  }\bibfield  {title} {\bibinfo {title} {Trans-spectral orbital angular
  momentum transfer via four-wave mixing in rb vapor},\ }\href@noop {}
  {\bibfield  {journal} {\bibinfo  {journal} {Physical review letters}\
  }\textbf {\bibinfo {volume} {108}},\ \bibinfo {pages} {243601} (\bibinfo
  {year} {2012})}\BibitemShut {NoStop}%
\bibitem [{\citenamefont {MATHEW}\ \emph {et~al.}(2021)\citenamefont {MATHEW}
  \emph {et~al.}}]{mathew2021single}%
  \BibitemOpen
  \bibfield  {author} {\bibinfo {author} {\bibfnamefont {R.}~\bibnamefont
  {MATHEW}} \emph {et~al.},\ }\emph {\bibinfo {title} {Single-photon generation
  via four-wave mixing in a thermal rubidium vapour at a high magnetic
  field}},\ \href@noop {} {Ph.D. thesis},\ \bibinfo  {school} {Durham
  University} (\bibinfo {year} {2021})\BibitemShut {NoStop}%
\bibitem [{\citenamefont {Hendrickson}\ \emph {et~al.}(2010)\citenamefont
  {Hendrickson}, \citenamefont {Lai}, \citenamefont {Pittman},\ and\
  \citenamefont {Franson}}]{hendrickson2010observation}%
  \BibitemOpen
  \bibfield  {author} {\bibinfo {author} {\bibfnamefont {S.}~\bibnamefont
  {Hendrickson}}, \bibinfo {author} {\bibfnamefont {M.}~\bibnamefont {Lai}},
  \bibinfo {author} {\bibfnamefont {T.}~\bibnamefont {Pittman}},\ and\ \bibinfo
  {author} {\bibfnamefont {J.}~\bibnamefont {Franson}},\ }\bibfield  {title}
  {\bibinfo {title} {Observation of two-photon absorption at low power levels
  using tapered optical fibers in rubidium vapor},\ }\href@noop {} {\bibfield
  {journal} {\bibinfo  {journal} {Physical review letters}\ }\textbf {\bibinfo
  {volume} {105}},\ \bibinfo {pages} {173602} (\bibinfo {year}
  {2010})}\BibitemShut {NoStop}%
\bibitem [{\citenamefont {Tabakaev}\ \emph {et~al.}(2022)\citenamefont
  {Tabakaev}, \citenamefont {Djorovi{\'c}}, \citenamefont {La~Volpe},
  \citenamefont {Gaulier}, \citenamefont {Ghosh}, \citenamefont {Bonacina},
  \citenamefont {Wolf}, \citenamefont {Zbinden},\ and\ \citenamefont
  {Thew}}]{tabakaev2022spatial}%
  \BibitemOpen
  \bibfield  {author} {\bibinfo {author} {\bibfnamefont {D.}~\bibnamefont
  {Tabakaev}}, \bibinfo {author} {\bibfnamefont {A.}~\bibnamefont
  {Djorovi{\'c}}}, \bibinfo {author} {\bibfnamefont {L.}~\bibnamefont
  {La~Volpe}}, \bibinfo {author} {\bibfnamefont {G.}~\bibnamefont {Gaulier}},
  \bibinfo {author} {\bibfnamefont {S.}~\bibnamefont {Ghosh}}, \bibinfo
  {author} {\bibfnamefont {L.}~\bibnamefont {Bonacina}}, \bibinfo {author}
  {\bibfnamefont {J.-P.}\ \bibnamefont {Wolf}}, \bibinfo {author}
  {\bibfnamefont {H.}~\bibnamefont {Zbinden}},\ and\ \bibinfo {author}
  {\bibfnamefont {R.}~\bibnamefont {Thew}},\ }\bibfield  {title} {\bibinfo
  {title} {Spatial properties of entangled two-photon absorption},\ }\href@noop
  {} {\bibfield  {journal} {\bibinfo  {journal} {Physical Review Letters}\
  }\textbf {\bibinfo {volume} {129}},\ \bibinfo {pages} {183601} (\bibinfo
  {year} {2022})}\BibitemShut {NoStop}%
\bibitem [{\citenamefont {Thaicharoen}\ \emph {et~al.}(2024)\citenamefont
  {Thaicharoen}, \citenamefont {Cardman},\ and\ \citenamefont
  {Raithel}}]{thaicharoen2024rydberg}%
  \BibitemOpen
  \bibfield  {author} {\bibinfo {author} {\bibfnamefont {N.}~\bibnamefont
  {Thaicharoen}}, \bibinfo {author} {\bibfnamefont {R.}~\bibnamefont
  {Cardman}},\ and\ \bibinfo {author} {\bibfnamefont {G.}~\bibnamefont
  {Raithel}},\ }\bibfield  {title} {\bibinfo {title} {Rydberg
  electromagnetically induced transparency of 85 rb vapor in a cell with ne
  buffer gas},\ }\href@noop {} {\bibfield  {journal} {\bibinfo  {journal}
  {Physical Review Applied}\ }\textbf {\bibinfo {volume} {21}},\ \bibinfo
  {pages} {064004} (\bibinfo {year} {2024})}\BibitemShut {NoStop}%
\bibitem [{\citenamefont {Pati}\ \emph {et~al.}(2021)\citenamefont {Pati},
  \citenamefont {Tripathi}, \citenamefont {Grewal}, \citenamefont {Pulido},\
  and\ \citenamefont {Depto}}]{pati2021synchronous}%
  \BibitemOpen
  \bibfield  {author} {\bibinfo {author} {\bibfnamefont {G.~S.}\ \bibnamefont
  {Pati}}, \bibinfo {author} {\bibfnamefont {R.}~\bibnamefont {Tripathi}},
  \bibinfo {author} {\bibfnamefont {R.~S.}\ \bibnamefont {Grewal}}, \bibinfo
  {author} {\bibfnamefont {M.}~\bibnamefont {Pulido}},\ and\ \bibinfo {author}
  {\bibfnamefont {R.~A.}\ \bibnamefont {Depto}},\ }\bibfield  {title} {\bibinfo
  {title} {Synchronous coherent population trapping and its magnetic spectral
  response in rubidium vapor},\ }\href@noop {} {\bibfield  {journal} {\bibinfo
  {journal} {Physical Review A}\ }\textbf {\bibinfo {volume} {104}},\ \bibinfo
  {pages} {033116} (\bibinfo {year} {2021})}\BibitemShut {NoStop}%
\bibitem [{\citenamefont {Brazhnikov}\ \emph {et~al.}(2019)\citenamefont
  {Brazhnikov}, \citenamefont {Ignatovich}, \citenamefont {Novokreshchenov},\
  and\ \citenamefont {Skvortsov}}]{brazhnikov2019ultrahigh}%
  \BibitemOpen
  \bibfield  {author} {\bibinfo {author} {\bibfnamefont {D.}~\bibnamefont
  {Brazhnikov}}, \bibinfo {author} {\bibfnamefont {S.}~\bibnamefont
  {Ignatovich}}, \bibinfo {author} {\bibfnamefont {A.}~\bibnamefont
  {Novokreshchenov}},\ and\ \bibinfo {author} {\bibfnamefont {M.}~\bibnamefont
  {Skvortsov}},\ }\bibfield  {title} {\bibinfo {title} {Ultrahigh-quality
  electromagnetically induced absorption resonances in a cesium vapor cell},\
  }\href@noop {} {\bibfield  {journal} {\bibinfo  {journal} {Journal of Physics
  B: Atomic, Molecular and Optical Physics}\ }\textbf {\bibinfo {volume}
  {52}},\ \bibinfo {pages} {215002} (\bibinfo {year} {2019})}\BibitemShut
  {NoStop}%
\bibitem [{\citenamefont {Ma}\ \emph {et~al.}(2022)\citenamefont {Ma},
  \citenamefont {Lei}, \citenamefont {Yan}, \citenamefont {Li}, \citenamefont
  {Chai}, \citenamefont {Yan}, \citenamefont {Jia}, \citenamefont {Xie},\ and\
  \citenamefont {Peng}}]{ma2022high}%
  \BibitemOpen
  \bibfield  {author} {\bibinfo {author} {\bibfnamefont {L.}~\bibnamefont
  {Ma}}, \bibinfo {author} {\bibfnamefont {X.}~\bibnamefont {Lei}}, \bibinfo
  {author} {\bibfnamefont {J.}~\bibnamefont {Yan}}, \bibinfo {author}
  {\bibfnamefont {R.}~\bibnamefont {Li}}, \bibinfo {author} {\bibfnamefont
  {T.}~\bibnamefont {Chai}}, \bibinfo {author} {\bibfnamefont {Z.}~\bibnamefont
  {Yan}}, \bibinfo {author} {\bibfnamefont {X.}~\bibnamefont {Jia}}, \bibinfo
  {author} {\bibfnamefont {C.}~\bibnamefont {Xie}},\ and\ \bibinfo {author}
  {\bibfnamefont {K.}~\bibnamefont {Peng}},\ }\bibfield  {title} {\bibinfo
  {title} {High-performance cavity-enhanced quantum memory with warm atomic
  cell},\ }\href@noop {} {\bibfield  {journal} {\bibinfo  {journal} {Nature
  communications}\ }\textbf {\bibinfo {volume} {13}},\ \bibinfo {pages} {2368}
  (\bibinfo {year} {2022})}\BibitemShut {NoStop}%
\bibitem [{\citenamefont {Lei}\ \emph {et~al.}(2022)\citenamefont {Lei},
  \citenamefont {Ma}, \citenamefont {Yan}, \citenamefont {Zhou}, \citenamefont
  {Yan},\ and\ \citenamefont {Jia}}]{lei2022electromagnetically}%
  \BibitemOpen
  \bibfield  {author} {\bibinfo {author} {\bibfnamefont {X.}~\bibnamefont
  {Lei}}, \bibinfo {author} {\bibfnamefont {L.}~\bibnamefont {Ma}}, \bibinfo
  {author} {\bibfnamefont {J.}~\bibnamefont {Yan}}, \bibinfo {author}
  {\bibfnamefont {X.}~\bibnamefont {Zhou}}, \bibinfo {author} {\bibfnamefont
  {Z.}~\bibnamefont {Yan}},\ and\ \bibinfo {author} {\bibfnamefont
  {X.}~\bibnamefont {Jia}},\ }\bibfield  {title} {\bibinfo {title}
  {Electromagnetically induced transparency quantum memory for non-classical
  states of light},\ }\href@noop {} {\bibfield  {journal} {\bibinfo  {journal}
  {Advances in Physics: X}\ }\textbf {\bibinfo {volume} {7}},\ \bibinfo {pages}
  {2060133} (\bibinfo {year} {2022})}\BibitemShut {NoStop}%
\bibitem [{\citenamefont {DeRose}\ \emph {et~al.}(2023)\citenamefont {DeRose},
  \citenamefont {Jiang}, \citenamefont {Li}, \citenamefont {Julius},
  \citenamefont {Zhuo}, \citenamefont {Wenner},\ and\ \citenamefont
  {Bali}}]{derose2023producing}%
  \BibitemOpen
  \bibfield  {author} {\bibinfo {author} {\bibfnamefont {K.}~\bibnamefont
  {DeRose}}, \bibinfo {author} {\bibfnamefont {K.}~\bibnamefont {Jiang}},
  \bibinfo {author} {\bibfnamefont {J.}~\bibnamefont {Li}}, \bibinfo {author}
  {\bibfnamefont {M.}~\bibnamefont {Julius}}, \bibinfo {author} {\bibfnamefont
  {L.}~\bibnamefont {Zhuo}}, \bibinfo {author} {\bibfnamefont {S.}~\bibnamefont
  {Wenner}},\ and\ \bibinfo {author} {\bibfnamefont {S.}~\bibnamefont {Bali}},\
  }\bibfield  {title} {\bibinfo {title} {Producing slow light in warm alkali
  vapor using electromagnetically induced transparency},\ }\href@noop {}
  {\bibfield  {journal} {\bibinfo  {journal} {American Journal of Physics}\
  }\textbf {\bibinfo {volume} {91}},\ \bibinfo {pages} {193} (\bibinfo {year}
  {2023})}\BibitemShut {NoStop}%
\bibitem [{\citenamefont {Chen}(2024)}]{chen2024triple}%
  \BibitemOpen
  \bibfield  {author} {\bibinfo {author} {\bibfnamefont {H.-J.}\ \bibnamefont
  {Chen}},\ }\bibfield  {title} {\bibinfo {title} {Triple electromagnetically
  induced transparency generated slow light for multiple carbon nanotube
  resonators},\ }\href@noop {} {\bibfield  {journal} {\bibinfo  {journal}
  {Journal of Applied Physics}\ }\textbf {\bibinfo {volume} {135}} (\bibinfo
  {year} {2024})}\BibitemShut {NoStop}%
\bibitem [{\citenamefont {Gonzalez~Maldonado}\ \emph
  {et~al.}(2024)\citenamefont {Gonzalez~Maldonado}, \citenamefont {Rollins},
  \citenamefont {Toyryla}, \citenamefont {McKelvy}, \citenamefont {Matsko},
  \citenamefont {Fan}, \citenamefont {Li}, \citenamefont {Wang}, \citenamefont
  {Kitching}, \citenamefont {Novikova} \emph
  {et~al.}}]{gonzalez2024sensitivity}%
  \BibitemOpen
  \bibfield  {author} {\bibinfo {author} {\bibfnamefont {M.}~\bibnamefont
  {Gonzalez~Maldonado}}, \bibinfo {author} {\bibfnamefont {O.}~\bibnamefont
  {Rollins}}, \bibinfo {author} {\bibfnamefont {A.}~\bibnamefont {Toyryla}},
  \bibinfo {author} {\bibfnamefont {J.~A.}\ \bibnamefont {McKelvy}}, \bibinfo
  {author} {\bibfnamefont {A.}~\bibnamefont {Matsko}}, \bibinfo {author}
  {\bibfnamefont {I.}~\bibnamefont {Fan}}, \bibinfo {author} {\bibfnamefont
  {Y.}~\bibnamefont {Li}}, \bibinfo {author} {\bibfnamefont {Y.-J.}\
  \bibnamefont {Wang}}, \bibinfo {author} {\bibfnamefont {J.}~\bibnamefont
  {Kitching}}, \bibinfo {author} {\bibfnamefont {I.}~\bibnamefont {Novikova}},
  \emph {et~al.},\ }\bibfield  {title} {\bibinfo {title} {Sensitivity of a
  vector atomic magnetometer based on electromagnetically induced
  transparency},\ }\href@noop {} {\bibfield  {journal} {\bibinfo  {journal}
  {Optics Express}\ }\textbf {\bibinfo {volume} {32}},\ \bibinfo {pages}
  {25062} (\bibinfo {year} {2024})}\BibitemShut {NoStop}%
\bibitem [{\citenamefont {Andryushkov}\ \emph {et~al.}(2022)\citenamefont
  {Andryushkov}, \citenamefont {Radnatarov},\ and\ \citenamefont
  {Kobtsev}}]{andryushkov2022vector}%
  \BibitemOpen
  \bibfield  {author} {\bibinfo {author} {\bibfnamefont {V.}~\bibnamefont
  {Andryushkov}}, \bibinfo {author} {\bibfnamefont {D.}~\bibnamefont
  {Radnatarov}},\ and\ \bibinfo {author} {\bibfnamefont {S.}~\bibnamefont
  {Kobtsev}},\ }\bibfield  {title} {\bibinfo {title} {Vector magnetometer based
  on the effect of coherent population trapping},\ }\href@noop {} {\bibfield
  {journal} {\bibinfo  {journal} {Applied Optics}\ }\textbf {\bibinfo {volume}
  {61}},\ \bibinfo {pages} {3604} (\bibinfo {year} {2022})}\BibitemShut
  {NoStop}%
\bibitem [{\citenamefont {Krysenko}\ \emph {et~al.}(2023)\citenamefont
  {Krysenko}, \citenamefont {Prudnikov}, \citenamefont {Taichenachev},
  \citenamefont {Yudin}, \citenamefont {Chepurov},\ and\ \citenamefont
  {Bagaev}}]{krysenko2023ground}%
  \BibitemOpen
  \bibfield  {author} {\bibinfo {author} {\bibfnamefont {D.}~\bibnamefont
  {Krysenko}}, \bibinfo {author} {\bibfnamefont {O.}~\bibnamefont {Prudnikov}},
  \bibinfo {author} {\bibfnamefont {A.}~\bibnamefont {Taichenachev}}, \bibinfo
  {author} {\bibfnamefont {V.}~\bibnamefont {Yudin}}, \bibinfo {author}
  {\bibfnamefont {S.}~\bibnamefont {Chepurov}},\ and\ \bibinfo {author}
  {\bibfnamefont {S.}~\bibnamefont {Bagaev}},\ }\bibfield  {title} {\bibinfo
  {title} {Ground-state electromagnetically-induced-transparency cooling of yb+
  171 ions in a polychromatic field},\ }\href@noop {} {\bibfield  {journal}
  {\bibinfo  {journal} {Physical Review A}\ }\textbf {\bibinfo {volume}
  {108}},\ \bibinfo {pages} {043114} (\bibinfo {year} {2023})}\BibitemShut
  {NoStop}%
\bibitem [{\citenamefont {Khazali}(2023)}]{khazali2023all}%
  \BibitemOpen
  \bibfield  {author} {\bibinfo {author} {\bibfnamefont {M.}~\bibnamefont
  {Khazali}},\ }\bibfield  {title} {\bibinfo {title} {All-optical quantum
  information processing via a single-step rydberg blockade gate},\ }\href@noop
  {} {\bibfield  {journal} {\bibinfo  {journal} {Optics Express}\ }\textbf
  {\bibinfo {volume} {31}},\ \bibinfo {pages} {13970} (\bibinfo {year}
  {2023})}\BibitemShut {NoStop}%
\bibitem [{\citenamefont {Shao}\ \emph {et~al.}(2024)\citenamefont {Shao},
  \citenamefont {Su}, \citenamefont {Li}, \citenamefont {Nath}, \citenamefont
  {Wu},\ and\ \citenamefont {Li}}]{shao2024rydberg}%
  \BibitemOpen
  \bibfield  {author} {\bibinfo {author} {\bibfnamefont {X.-Q.}\ \bibnamefont
  {Shao}}, \bibinfo {author} {\bibfnamefont {S.-L.}\ \bibnamefont {Su}},
  \bibinfo {author} {\bibfnamefont {L.}~\bibnamefont {Li}}, \bibinfo {author}
  {\bibfnamefont {R.}~\bibnamefont {Nath}}, \bibinfo {author} {\bibfnamefont
  {J.-H.}\ \bibnamefont {Wu}},\ and\ \bibinfo {author} {\bibfnamefont
  {W.}~\bibnamefont {Li}},\ }\bibfield  {title} {\bibinfo {title} {Rydberg
  superatoms: An artificial quantum system for quantum information processing
  and quantum optics},\ }\href@noop {} {\bibfield  {journal} {\bibinfo
  {journal} {Applied Physics Reviews}\ }\textbf {\bibinfo {volume} {11}}
  (\bibinfo {year} {2024})}\BibitemShut {NoStop}%
\bibitem [{\citenamefont {Chanu}\ \emph {et~al.}(2012)\citenamefont {Chanu},
  \citenamefont {Pandey},\ and\ \citenamefont
  {Natarajan}}]{chanu2012conversion}%
  \BibitemOpen
  \bibfield  {author} {\bibinfo {author} {\bibfnamefont {S.~R.}\ \bibnamefont
  {Chanu}}, \bibinfo {author} {\bibfnamefont {K.}~\bibnamefont {Pandey}},\ and\
  \bibinfo {author} {\bibfnamefont {V.}~\bibnamefont {Natarajan}},\ }\bibfield
  {title} {\bibinfo {title} {Conversion between electromagnetically induced
  transparency and absorption in a three-level lambda system},\ }\href@noop {}
  {\bibfield  {journal} {\bibinfo  {journal} {Europhysics Letters}\ }\textbf
  {\bibinfo {volume} {98}},\ \bibinfo {pages} {44009} (\bibinfo {year}
  {2012})}\BibitemShut {NoStop}%
\bibitem [{\citenamefont {Singh}\ \emph {et~al.}(2023)\citenamefont {Singh},
  \citenamefont {Sharma}, \citenamefont {Subba}, \citenamefont {Chatterjee},\
  and\ \citenamefont {Tripathi}}]{singh2023light}%
  \BibitemOpen
  \bibfield  {author} {\bibinfo {author} {\bibfnamefont {R.~K.}\ \bibnamefont
  {Singh}}, \bibinfo {author} {\bibfnamefont {N.}~\bibnamefont {Sharma}},
  \bibinfo {author} {\bibfnamefont {I.~H.}\ \bibnamefont {Subba}}, \bibinfo
  {author} {\bibfnamefont {S.}~\bibnamefont {Chatterjee}},\ and\ \bibinfo
  {author} {\bibfnamefont {A.}~\bibnamefont {Tripathi}},\ }\bibfield  {title}
  {\bibinfo {title} {Light shift induced modification of electromagnetically
  induced resonances in atomic vapor},\ }\href@noop {} {\bibfield  {journal}
  {\bibinfo  {journal} {Optics Communications}\ }\textbf {\bibinfo {volume}
  {537}},\ \bibinfo {pages} {129466} (\bibinfo {year} {2023})}\BibitemShut
  {NoStop}%
\bibitem [{\citenamefont {Xu}\ \emph {et~al.}(2013)\citenamefont {Xu},
  \citenamefont {Shen},\ and\ \citenamefont {Xiao}}]{xu2013tuning}%
  \BibitemOpen
  \bibfield  {author} {\bibinfo {author} {\bibfnamefont {X.}~\bibnamefont
  {Xu}}, \bibinfo {author} {\bibfnamefont {S.}~\bibnamefont {Shen}},\ and\
  \bibinfo {author} {\bibfnamefont {Y.}~\bibnamefont {Xiao}},\ }\bibfield
  {title} {\bibinfo {title} {Tuning the phase sensitivity of a double-lambda
  system with a static magnetic field},\ }\href@noop {} {\bibfield  {journal}
  {\bibinfo  {journal} {Optics Express}\ }\textbf {\bibinfo {volume} {21}},\
  \bibinfo {pages} {11705} (\bibinfo {year} {2013})}\BibitemShut {NoStop}%
\bibitem [{\citenamefont {Zhang}\ \emph {et~al.}(2024)\citenamefont {Zhang},
  \citenamefont {Kong},\ and\ \citenamefont {Wang}}]{zhang2024controlling}%
  \BibitemOpen
  \bibfield  {author} {\bibinfo {author} {\bibfnamefont {Y.}~\bibnamefont
  {Zhang}}, \bibinfo {author} {\bibfnamefont {D.}~\bibnamefont {Kong}},\ and\
  \bibinfo {author} {\bibfnamefont {F.}~\bibnamefont {Wang}},\ }\bibfield
  {title} {\bibinfo {title} {Controlling of the spatially dependent absorption
  spectrum by quantum interference},\ }\href@noop {} {\bibfield  {journal}
  {\bibinfo  {journal} {Applied Optics}\ }\textbf {\bibinfo {volume} {63}},\
  \bibinfo {pages} {7410} (\bibinfo {year} {2024})}\BibitemShut {NoStop}%
\bibitem [{\citenamefont {Bhattarai}\ \emph {et~al.}(2018)\citenamefont
  {Bhattarai}, \citenamefont {Bharti},\ and\ \citenamefont
  {Natarajan}}]{bhattarai2018tuning}%
  \BibitemOpen
  \bibfield  {author} {\bibinfo {author} {\bibfnamefont {M.}~\bibnamefont
  {Bhattarai}}, \bibinfo {author} {\bibfnamefont {V.}~\bibnamefont {Bharti}},\
  and\ \bibinfo {author} {\bibfnamefont {V.}~\bibnamefont {Natarajan}},\
  }\bibfield  {title} {\bibinfo {title} {Tuning of the hanle effect from eit to
  eia using spatially separated probe and control beams},\ }\href@noop {}
  {\bibfield  {journal} {\bibinfo  {journal} {Scientific Reports}\ }\textbf
  {\bibinfo {volume} {8}},\ \bibinfo {pages} {7525} (\bibinfo {year}
  {2018})}\BibitemShut {NoStop}%
\bibitem [{\citenamefont {Ram}\ \emph {et~al.}(2010)\citenamefont {Ram},
  \citenamefont {Pattabiraman},\ and\ \citenamefont {Vijayan}}]{ram2010effect}%
  \BibitemOpen
  \bibfield  {author} {\bibinfo {author} {\bibfnamefont {N.}~\bibnamefont
  {Ram}}, \bibinfo {author} {\bibfnamefont {M.}~\bibnamefont {Pattabiraman}},\
  and\ \bibinfo {author} {\bibfnamefont {C.}~\bibnamefont {Vijayan}},\
  }\bibfield  {title} {\bibinfo {title} {Effect of ellipticity on hanle
  electromagnetically induced absorption and transparency resonances with
  longitudinal and transverse magnetic fields},\ }\href@noop {} {\bibfield
  {journal} {\bibinfo  {journal} {Physical Review A}\ }\textbf {\bibinfo
  {volume} {82}},\ \bibinfo {pages} {033417} (\bibinfo {year}
  {2010})}\BibitemShut {NoStop}%
\bibitem [{\citenamefont {Brazhnikov}\ \emph {et~al.}(2011)\citenamefont
  {Brazhnikov}, \citenamefont {Taichenachev},\ and\ \citenamefont
  {Yudin}}]{brazhnikov2011polarization}%
  \BibitemOpen
  \bibfield  {author} {\bibinfo {author} {\bibfnamefont {D.~V.}\ \bibnamefont
  {Brazhnikov}}, \bibinfo {author} {\bibfnamefont {A.~V.}\ \bibnamefont
  {Taichenachev}},\ and\ \bibinfo {author} {\bibfnamefont {V.~I.}\ \bibnamefont
  {Yudin}},\ }\bibfield  {title} {\bibinfo {title} {Polarization method for
  controlling a sign of electromagnetically-induced transparency/absorption
  resonances},\ }\href@noop {} {\bibfield  {journal} {\bibinfo  {journal} {The
  European Physical Journal D}\ }\textbf {\bibinfo {volume} {63}},\ \bibinfo
  {pages} {315} (\bibinfo {year} {2011})}\BibitemShut {NoStop}%
\bibitem [{\citenamefont {Chauhan}\ \emph {et~al.}(2021)\citenamefont
  {Chauhan}, \citenamefont {Kumar}, \citenamefont {Manchaiah},\ and\
  \citenamefont {Easwaran}}]{chauhan2021enhancement}%
  \BibitemOpen
  \bibfield  {author} {\bibinfo {author} {\bibfnamefont {V.~S.}\ \bibnamefont
  {Chauhan}}, \bibinfo {author} {\bibfnamefont {R.}~\bibnamefont {Kumar}},
  \bibinfo {author} {\bibfnamefont {D.}~\bibnamefont {Manchaiah}},\ and\
  \bibinfo {author} {\bibfnamefont {R.~K.}\ \bibnamefont {Easwaran}},\
  }\bibfield  {title} {\bibinfo {title} {Enhancement of electromagnetically
  induced transparency and absorption signals in 85 rb atomic vapor medium by
  using a small external magnetic field},\ }\href@noop {} {\bibfield  {journal}
  {\bibinfo  {journal} {JOSA B}\ }\textbf {\bibinfo {volume} {38}},\ \bibinfo
  {pages} {630} (\bibinfo {year} {2021})}\BibitemShut {NoStop}%
\bibitem [{\citenamefont {Sargsyan}\ \emph {et~al.}(2012)\citenamefont
  {Sargsyan}, \citenamefont {Mirzoyan},\ and\ \citenamefont
  {Sarkisyan}}]{sargsyan2012splitting}%
  \BibitemOpen
  \bibfield  {author} {\bibinfo {author} {\bibfnamefont {A.}~\bibnamefont
  {Sargsyan}}, \bibinfo {author} {\bibfnamefont {R.}~\bibnamefont {Mirzoyan}},\
  and\ \bibinfo {author} {\bibfnamefont {D.}~\bibnamefont {Sarkisyan}},\
  }\bibfield  {title} {\bibinfo {title} {Splitting of the electromagnetically
  induced transparency resonance on 85 rb atoms in strong magnetic fields up to
  the paschen-back regime},\ }\href@noop {} {\bibfield  {journal} {\bibinfo
  {journal} {JETP letters}\ }\textbf {\bibinfo {volume} {96}},\ \bibinfo
  {pages} {303} (\bibinfo {year} {2012})}\BibitemShut {NoStop}%
\bibitem [{\citenamefont {Mishra}\ \emph {et~al.}(2018)\citenamefont {Mishra},
  \citenamefont {Chakraborty}, \citenamefont {Srivastava}, \citenamefont
  {Tiwari}, \citenamefont {Ram}, \citenamefont {Tiwari},\ and\ \citenamefont
  {Mishra}}]{mishra2018electromagnetically}%
  \BibitemOpen
  \bibfield  {author} {\bibinfo {author} {\bibfnamefont {C.}~\bibnamefont
  {Mishra}}, \bibinfo {author} {\bibfnamefont {A.}~\bibnamefont {Chakraborty}},
  \bibinfo {author} {\bibfnamefont {A.}~\bibnamefont {Srivastava}}, \bibinfo
  {author} {\bibfnamefont {S.}~\bibnamefont {Tiwari}}, \bibinfo {author}
  {\bibfnamefont {S.}~\bibnamefont {Ram}}, \bibinfo {author} {\bibfnamefont
  {V.}~\bibnamefont {Tiwari}},\ and\ \bibinfo {author} {\bibfnamefont
  {S.}~\bibnamefont {Mishra}},\ }\bibfield  {title} {\bibinfo {title}
  {Electromagnetically induced transparency in $\lambda$-systems of 87 rb atom
  in magnetic field},\ }\href@noop {} {\bibfield  {journal} {\bibinfo
  {journal} {Journal of Modern Optics}\ }\textbf {\bibinfo {volume} {65}},\
  \bibinfo {pages} {2269} (\bibinfo {year} {2018})}\BibitemShut {NoStop}%
\bibitem [{\citenamefont {Bhushan}\ \emph {et~al.}(2019)\citenamefont
  {Bhushan}, \citenamefont {Chauhan}, \citenamefont {Dixith},\ and\
  \citenamefont {Easwaran}}]{bhushan2019effect}%
  \BibitemOpen
  \bibfield  {author} {\bibinfo {author} {\bibfnamefont {S.}~\bibnamefont
  {Bhushan}}, \bibinfo {author} {\bibfnamefont {V.~S.}\ \bibnamefont
  {Chauhan}}, \bibinfo {author} {\bibfnamefont {M.}~\bibnamefont {Dixith}},\
  and\ \bibinfo {author} {\bibfnamefont {R.~K.}\ \bibnamefont {Easwaran}},\
  }\bibfield  {title} {\bibinfo {title} {Effect of magnetic field on a multi
  window ladder type electromagnetically induced transparency with 87rb atoms
  in vapour cell},\ }\href@noop {} {\bibfield  {journal} {\bibinfo  {journal}
  {Physics Letters A}\ }\textbf {\bibinfo {volume} {383}},\ \bibinfo {pages}
  {125885} (\bibinfo {year} {2019})}\BibitemShut {NoStop}%
\bibitem [{\citenamefont {Ma}\ \emph {et~al.}(2017)\citenamefont {Ma},
  \citenamefont {Anderson},\ and\ \citenamefont {Raithel}}]{ma2017paschen}%
  \BibitemOpen
  \bibfield  {author} {\bibinfo {author} {\bibfnamefont {L.}~\bibnamefont
  {Ma}}, \bibinfo {author} {\bibfnamefont {D.}~\bibnamefont {Anderson}},\ and\
  \bibinfo {author} {\bibfnamefont {G.}~\bibnamefont {Raithel}},\ }\bibfield
  {title} {\bibinfo {title} {Paschen-back effects and rydberg-state
  diamagnetism in vapor-cell electromagnetically induced transparency},\
  }\href@noop {} {\bibfield  {journal} {\bibinfo  {journal} {Physical Review
  A}\ }\textbf {\bibinfo {volume} {95}},\ \bibinfo {pages} {061804} (\bibinfo
  {year} {2017})}\BibitemShut {NoStop}%
\bibitem [{\citenamefont {Singh}\ \emph {et~al.}(2021)\citenamefont {Singh},
  \citenamefont {Sharma}, \citenamefont {Subba}, \citenamefont {Chatterjee},\
  and\ \citenamefont {Tripathi}}]{singh2021competition}%
  \BibitemOpen
  \bibfield  {author} {\bibinfo {author} {\bibfnamefont {R.~K.}\ \bibnamefont
  {Singh}}, \bibinfo {author} {\bibfnamefont {N.}~\bibnamefont {Sharma}},
  \bibinfo {author} {\bibfnamefont {I.~H.}\ \bibnamefont {Subba}}, \bibinfo
  {author} {\bibfnamefont {S.}~\bibnamefont {Chatterjee}},\ and\ \bibinfo
  {author} {\bibfnamefont {A.}~\bibnamefont {Tripathi}},\ }\bibfield  {title}
  {\bibinfo {title} {Competition between off-resonant and on-resonant processes
  in electromagnetically induced transparency in presence of magnetic field},\
  }\href@noop {} {\bibfield  {journal} {\bibinfo  {journal} {Physics Letters
  A}\ }\textbf {\bibinfo {volume} {416}},\ \bibinfo {pages} {127673} (\bibinfo
  {year} {2021})}\BibitemShut {NoStop}%
\bibitem [{\citenamefont {Subba}\ \emph {et~al.}(2020)\citenamefont {Subba},
  \citenamefont {Singh}, \citenamefont {Sharma}, \citenamefont {Chatterjee},\
  and\ \citenamefont {Tripathi}}]{subba2020understanding}%
  \BibitemOpen
  \bibfield  {author} {\bibinfo {author} {\bibfnamefont {I.~H.}\ \bibnamefont
  {Subba}}, \bibinfo {author} {\bibfnamefont {R.~K.}\ \bibnamefont {Singh}},
  \bibinfo {author} {\bibfnamefont {N.}~\bibnamefont {Sharma}}, \bibinfo
  {author} {\bibfnamefont {S.}~\bibnamefont {Chatterjee}},\ and\ \bibinfo
  {author} {\bibfnamefont {A.}~\bibnamefont {Tripathi}},\ }\bibfield  {title}
  {\bibinfo {title} {Understanding asymmetry in electromagnetically induced
  transparency for 87 rb in strong transverse magnetic field},\ }\href@noop {}
  {\bibfield  {journal} {\bibinfo  {journal} {The European Physical Journal D}\
  }\textbf {\bibinfo {volume} {74}},\ \bibinfo {pages} {1} (\bibinfo {year}
  {2020})}\BibitemShut {NoStop}%
\bibitem [{\citenamefont {Das}\ \emph {et~al.}(2021)\citenamefont {Das},
  \citenamefont {Das}, \citenamefont {Bhattacharyya},\ and\ \citenamefont
  {De}}]{das2021effects}%
  \BibitemOpen
  \bibfield  {author} {\bibinfo {author} {\bibfnamefont {B.~C.}\ \bibnamefont
  {Das}}, \bibinfo {author} {\bibfnamefont {A.}~\bibnamefont {Das}}, \bibinfo
  {author} {\bibfnamefont {D.}~\bibnamefont {Bhattacharyya}},\ and\ \bibinfo
  {author} {\bibfnamefont {S.}~\bibnamefont {De}},\ }\bibfield  {title}
  {\bibinfo {title} {Effects of vector magnetic field on electromagnetically
  induced transparency with lin perp. lin polarization},\ }\href@noop {}
  {\bibfield  {journal} {\bibinfo  {journal} {JOSA B}\ }\textbf {\bibinfo
  {volume} {38}},\ \bibinfo {pages} {584} (\bibinfo {year} {2021})}\BibitemShut
  {NoStop}%
\bibitem [{\citenamefont {Bhattarai}\ \emph {et~al.}(2019)\citenamefont
  {Bhattarai}, \citenamefont {Bharti}, \citenamefont {Natarajan}, \citenamefont
  {Sargsyan},\ and\ \citenamefont {Sarkisyan}}]{bhattarai2019study}%
  \BibitemOpen
  \bibfield  {author} {\bibinfo {author} {\bibfnamefont {M.}~\bibnamefont
  {Bhattarai}}, \bibinfo {author} {\bibfnamefont {V.}~\bibnamefont {Bharti}},
  \bibinfo {author} {\bibfnamefont {V.}~\bibnamefont {Natarajan}}, \bibinfo
  {author} {\bibfnamefont {A.}~\bibnamefont {Sargsyan}},\ and\ \bibinfo
  {author} {\bibfnamefont {D.}~\bibnamefont {Sarkisyan}},\ }\bibfield  {title}
  {\bibinfo {title} {Study of eit resonances in an anti-relaxation coated rb
  vapor cell},\ }\href@noop {} {\bibfield  {journal} {\bibinfo  {journal}
  {Physics Letters A}\ }\textbf {\bibinfo {volume} {383}},\ \bibinfo {pages}
  {91} (\bibinfo {year} {2019})}\BibitemShut {NoStop}%
\bibitem [{\citenamefont {Hassan}\ \emph {et~al.}(2025)\citenamefont {Hassan},
  \citenamefont {Noh},\ and\ \citenamefont {Kim}}]{hassan2025effect}%
  \BibitemOpen
  \bibfield  {author} {\bibinfo {author} {\bibfnamefont {A.-u.}\ \bibnamefont
  {Hassan}}, \bibinfo {author} {\bibfnamefont {H.-R.}\ \bibnamefont {Noh}},\
  and\ \bibinfo {author} {\bibfnamefont {J.-T.}\ \bibnamefont {Kim}},\
  }\bibfield  {title} {\bibinfo {title} {Effect of neighbouring transitions on
  critical angle in conversion between eia and eit in 87rb atoms},\ }\href@noop
  {} {\bibfield  {journal} {\bibinfo  {journal} {Journal of Modern Optics}\ ,\
  \bibinfo {pages} {1}} (\bibinfo {year} {2025})}\BibitemShut {NoStop}%
\bibitem [{\citenamefont {Milner}\ and\ \citenamefont
  {Prior}(1998)}]{milner1998multilevel}%
  \BibitemOpen
  \bibfield  {author} {\bibinfo {author} {\bibfnamefont {V.}~\bibnamefont
  {Milner}}\ and\ \bibinfo {author} {\bibfnamefont {Y.}~\bibnamefont {Prior}},\
  }\bibfield  {title} {\bibinfo {title} {Multilevel dark states: Coherent
  population trapping with elliptically polarized incoherent light},\
  }\href@noop {} {\bibfield  {journal} {\bibinfo  {journal} {Physical review
  letters}\ }\textbf {\bibinfo {volume} {80}},\ \bibinfo {pages} {940}
  (\bibinfo {year} {1998})}\BibitemShut {NoStop}%
\bibitem [{\citenamefont {Milner}\ \emph {et~al.}(1999)\citenamefont {Milner},
  \citenamefont {Chernobrod},\ and\ \citenamefont
  {Prior}}]{milner1999arbitrary}%
  \BibitemOpen
  \bibfield  {author} {\bibinfo {author} {\bibfnamefont {V.}~\bibnamefont
  {Milner}}, \bibinfo {author} {\bibfnamefont {B.~M.}\ \bibnamefont
  {Chernobrod}},\ and\ \bibinfo {author} {\bibfnamefont {Y.}~\bibnamefont
  {Prior}},\ }\bibfield  {title} {\bibinfo {title} {Arbitrary orientation of
  atoms and molecules via coherent population trapping by elliptically
  polarized light},\ }\href@noop {} {\bibfield  {journal} {\bibinfo  {journal}
  {Physical Review A}\ }\textbf {\bibinfo {volume} {60}},\ \bibinfo {pages}
  {1293} (\bibinfo {year} {1999})}\BibitemShut {NoStop}%
\bibitem [{\citenamefont {Steck}(2001)}]{steck2001rubidium}%
  \BibitemOpen
  \bibfield  {author} {\bibinfo {author} {\bibfnamefont {D.~A.}\ \bibnamefont
  {Steck}},\ }\bibfield  {title} {\bibinfo {title} {Rubidium 87 d line data},\
  }\href@noop {} {\  (\bibinfo {year} {2001})}\BibitemShut {NoStop}%
\end{thebibliography}%
\end{document}